\documentclass{emulateapj}
\usepackage{epsfig, xspace}

\newcommand{\saxj}{\mbox{SAX J1808.4$-$3658}\xspace}
\newcommand{\igrj}{\mbox{IGR J00291$+$5934}\xspace}
\newcommand{\RXTE}{\textit{RXTE}\xspace}
\newcommand{\Chandra}{\textit{Chandra}\xspace}
\newcommand{\XMM}{\textit{XMM}\xspace}
\newcommand{\INTEGRAL}{\textit{INTEGRAL}\xspace}
\newcommand{\Swift}{\textit{Swift}\xspace}
\newcommand{\Fermi}{\textit{Fermi}\xspace}
\newcommand{\gammaray}{\mbox{$\gamma$-ray}\xspace}

\newcommand{\Halpha}{H$\alpha$\xspace}
\newcommand{\us}{$\mu$s\xspace}
\newcommand{\uHz}{$\mu$Hz\xspace}
\newcommand{\HzPerSec}{~Hz~s$^{-1}$\xspace}
\newcommand{\fluxunits}{~erg~cm$^{-2}$~s$^{-1}$\xspace}

\newcommand{\sig}{$\,\sigma$\xspace}
\newcommand{\pz}{\phantom{0}}
\newcommand{\eq}[1]{eq.~(\ref{eq:#1})}

\begin{document}

\title{A Double Outburst from IGR~J00291$+$5934:\\
  Implications for Accretion Disk Instability Theory}
\shorttitle{Double Outburst from \igrj}
\shortauthors{Hartman et~al.}

\submitted{Accepted by ApJ}

\author{Jacob M. Hartman\altaffilmark{1,2},
Duncan K. Galloway\altaffilmark{3}, and Deepto Chakrabarty\altaffilmark{4}.}
\altaffiltext{1}{Code 7655, Naval Research Laboratory, Washington, DC 20375;
Jacob.Hartman@nrl.navy.mil}
\altaffiltext{2}{National Research Council research associate}
\altaffiltext{3}{School of Physics and School of Mathematical Sciences, Monash
University, Victoria 3800, Australia}
\altaffiltext{4}{Kavli Institute for Astrophysics and Space Research,
Massachusetts Institute of Technology, Cambridge, MA 02139}

\begin{abstract}

The accretion-powered millisecond pulsar \igrj underwent two $\sim$10~d long
outbursts during 2008, separated by 30~d in quiescence.  Such a short
quiescent period between outbursts has never been seen before from a neutron
star X-ray transient.  X-ray pulsations at the 599~Hz spin frequency are
detected throughout both outbursts.  For the first time, we derive a pulse
phase model that connects two outbursts, providing a long baseline for spin
frequency measurement.  Comparison with the frequency measured during the 2004
outburst of this source gives a spin-down during quiescence of
$-(4\pm1)\times10^{-15}$\HzPerSec, approximately an order of magnitude larger
than the long-term spin-down observed in the 401~Hz accretion-powered pulsar
\saxj.  If this spin-down is due to magnetic dipole radiation, it requires a
$2\times10^8$~G field strength, and its high spin-down luminosity may be
detectable with the \Fermi Large Area Telescope.  Alternatively, this large
spin-down could be produced by gravitational wave emission from a fractional
mass quadrupole moment of $Q/I = 1\times10^{-9}$.  The rapid succession of the
outbursts also provides a unique test of models for accretion in low-mass
X-ray binaries.  Disk instability models generally predict that an outburst
will leave the accretion disk too depleted to fuel a second outburst after
such a brief quiescence.  We suggest a modification in which the outburst is
shut off by the onset of a propeller effect before the disk is depleted.  This
model can explain the short quiescence and the unusually slow rise of the
light curve of the second 2008 outburst.

\end{abstract}
\keywords{binaries: general --- stars: individual (\igrj) --- stars: neutron
--- stars: rotation --- X-rays: binaries --- X-rays: stars}

\section{Introduction}

The longevity of the {\em Rossi X-ray Timing Explorer} (\RXTE) has provided
the opportunity to observe multiple outbursts from accretion-powered
millisecond pulsars (AMSPs) with recurrence times of $\sim$10~yr or less.
Such observations can address a diverse array of science.  The pulse timing of
multiple outbursts can measure changes in the spin of the neutron star (NS),
which places limits on its magnetic field and gravitational wave emission.
Comparison of the light curves and spin frequency derivatives during outburst
probes the interaction between the NS and the accretion disk, while the shape
of the pulses constrains the nature of the X-ray emission and the magnetically
channeled accretion flow.  Finally, the observation of multiple outbursts
provides tests of disk instability models and other theories put forward to
explain the recurrence of these sources.  In this paper, we report on the
observation by \RXTE of a second and third outburst of the 599~Hz accreting
pulsar \igrj.  These outbursts were separated by only 30~d of quiescence, a
more rapid recurrence than has ever been seen before from a NS low-mass X-ray
binary (LMXB).  The proximity of the two outbursts proves particularly useful
for addressing many of the above questions.

The {\em International Gamma-ray Astrophysics Labratory} (\INTEGRAL) first
detected \igrj at the onset of an outburst on 2004 December~2
\citep{Eckert04}.  Follow-up observations with \RXTE revealed pulsations at a
frequency of 598.89~Hz \citep{Markwardt04a}, modulated by a 147.4~minute orbit
\citep{Markwardt04b}.  Analysis of the pulsations revealed them to be highly
sinusoidal across a wide range of energies \citep{Galloway05, Falanga05}.
Pulses arrived progressively sooner with increasing energy over 2--8~keV,
following the pattern of soft lags observed in other AMSPs \citep{Galloway05},
but above 8~keV these soft lags diminished, a reversal not seen in other AMSPs
that may have important implications for the origin of these lags
\citep{Falanga07}.  Fractional amplitudes were between 5\%--10\% rms,
generally decreasing with energy \citep{Galloway05}.  Pulse timing models for
the 2004 outburst require a spin derivative of
$8.5(1.1)\times10^{-13}$\HzPerSec \citep{Falanga05, Burderi07}, which has been
ascribed to the NS being spun up by the accreting matter \citep{Burderi07}.
(Note that parenthetical uncertaintes are at the 1\sig level throughout this
paper.)  Aperiodic timing of \igrj reveals an unusual amount of timing noise
at very low frequencies (0.01--0.1~Hz), producing a timing spectrum more akin
to black holes than to other NS low-mass X-ray binaries \citep{Linares07}.

A mission-long light curve from the \RXTE All Sky Monitor showed marginal
(5\sig) detections of earlier outbursts during 1998 November and 2001
September \citep{Remillard04}.  With each outburst, the duration of quiescence
increased by 160--170~d.  A quadratic fit to these outburst times predicted a
3.6~yr quiescence between the 2004 outburst and the start of the 2008
outbursts \citep{Galloway08}.  This estimate proved to be accurate to within
1\% of the recurrence period, with the source returning to outburst on 2008
Aug 13 \citep{Chakrabarty08}.

X-ray observations between the 2008 outbursts show that \igrj reached a flux
level consistent with prior measurements during longer intervals of
quiescence.  On Aug~21 the \Swift X-ray Telescope (XRT) gave a 3\sig upper
limit on the unabsorbed 2--10~keV flux of $4.7\times10^{-12}$\fluxunits, and
on Aug~25 an \XMM observation detected the source at an unabsorbed 2--10~keV
flux of $(1.4\pm0.3)\times10^{-14}$\fluxunits \citep{Lewis10}.  In comparison,
a Chandra observation nearly two years after the 2004 outburst showed an
unabsorbed 0.5--10~keV flux of $1\times10^{-13}$\fluxunits \citep{Jonker08}.
Because the accretion episodes during 2008 were separated by a quiescent
period during which accretion was very low or halted entirely, we refer to
them as two separate outbursts, 2008a and 2008b.

In this paper, we present a detailed analysis of the 2008 outbursts of \igrj
and the implications for our understanding of recurrent X-ray transients.  We
describe the RXTE observations and data analysis in \S2, and we present the
results of this analysis in \S3.  In \S4 we examine what the 2008 double
outburst can tell us about the theories of accretion disk instability.  In \S5
we consider other evidence for changes in the accretion regime at low fluxes,
and in \S6 we derive limits on the NS magnetic field.  In \S7 we discuss the
long-term spin evolution and the possible sources of torque on the NS.  In a
final section, we summarize our results and conclusions.

\section{Observations and Analysis}

\begin{figure}[t!]
  \begin{center}
    \includegraphics[width=0.45\textwidth]{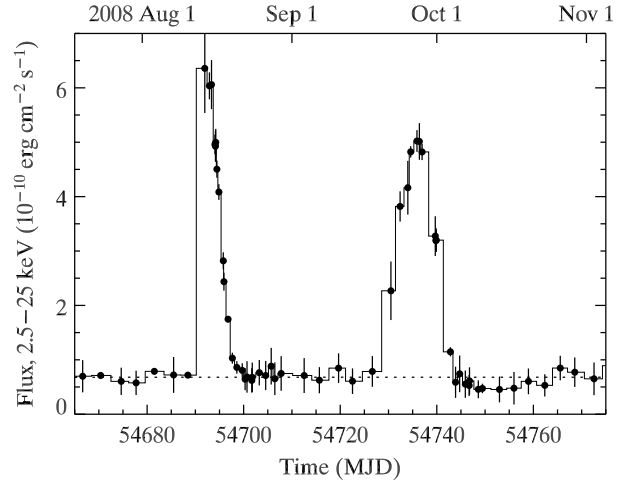}
  \end{center}
  \caption{ The \RXTE PCA light curve of the 2008 outbursts from \igrj.  The
    horizontal dotted line shows the estimated contribution from the nearby
    intermediate polar V709~Cas.
  \label{fig:Lightcurve}}
\end{figure}

\igrj was found to be in outburst during an \RXTE monitoring observation on
2008 Aug 13 \citep{Chakrabarty08}.  \RXTE observed the source daily during
this outburst (observation IDs \mbox{93013-07-*}).  By Aug 22, the source had
faded below the \RXTE detection threshold ($\sim$$1\times10^{-11}$\fluxunits).
Intensive \RXTE observation ended on Aug 30, and the program of twice-weekly
$\approx$1~ks monitoring observations (observation IDs \mbox{93435-01-*})
resumed.  On Sep 18, an optical observation found the source to have
unexpectedly re-brightened, and a \Swift XRT observation on Sep 20 confirmed
that \igrj was again in outburst \citep{Lewis08}.  Longer, more frequent \RXTE
observations resumed shortly thereafter.  This second outburst persisted
through Oct 3.  During the first outburst, 20 \RXTE observations were taken,
totaling 75.3~ks of exposure; during the second, 12 observations were taken,
totaling 37.9~ks.  Figure~\ref{fig:Lightcurve} shows the \RXTE light curve of
these outbursts and the times of observations.

We analyzed the data from the \RXTE Proportional Counter Array (PCA;
\citealt{Jahoda96}), which comprises five identical, co-aligned proportional
counter units (PCUs) sensitive to 2.5--60~keV photons within
$\approx$1\degr\ of the pointing axis.  Due to the increased frequency of
high-voltage breakdowns in the PCUs \citep{Jahoda06}, an average of only 1.7
PCUs were taking usable data during the observations.  (For comparison, an
average of 2.8 PCUs were active during the 2004 outburst of \igrj.)  The
resulting mean effective area during the 2008 outbursts was 2100~cm$^2$.

We derived outburst light curves using only PCU~2, which was active throughout
all the observations.  We fit each observation with an absorbed blackbody plus
power law model using XSPEC and the latest PCA response matrix (version 11.7)
to estimate the 2.5--25~keV flux and its uncertainty.  Note that the absorbed
fluxes are given throughout this paper (i.e., the flux as observed by \RXTE).
Instrumental background levels were estimated using the FTOOL {\tt
  pcabackest}.\footnote{\url{http://heasarc.gsfc.nasa.gov/docs/xte/recipes/pcabackest.html}}

The large field of view of the PCA also admits photons from the nearby
intermediate polar V709~Cas, located 17 arcmin from \igrj.  For all
observations the \RXTE science axis was pointed at \igrj, resulting in the PCA
collimator admitting 70\% of the photons from V709~Cas (using the linear
collimator model from \S8.3 of \citealt{Jahoda06}).  We estimated its
contribution using the monitoring observations of the \igrj field that
preceded and followed the 2008 outbursts.  The mean 2.5--25~keV flux was
$0.68\times10^{-10}$\fluxunits after subtracting the instrumental background,
implying an unattenuated flux from V709~Cas of $1.0\times10^{-10}$\fluxunits.
Earlier observations suggest that its 2.5--25~keV flux varies within
$(0.9$--$1.3)\times10^{-10}$\fluxunits (\citealt{DeMartino01, Falanga05b}; and
references therein), consistent with our results.  The uncertainty in our
V709~Cas flux measurements sets a PCA detection threshold of
$\sim$$1\times10^{-11}$\fluxunits for \igrj.  The flux of V709~Cas is
modulated by its 312.8~s rotational period \citep{Haberl95} with a fractional
amplitude of 20--30\% \citep{DeMartino01}.  When dividing \RXTE observations
into shorter intervals, we chose integral multiples of the V709~Cas period to
simplify the estimation of its contribution.

\begin{figure*}[t!]
  \begin{center}
    \includegraphics[width=0.85\textwidth]{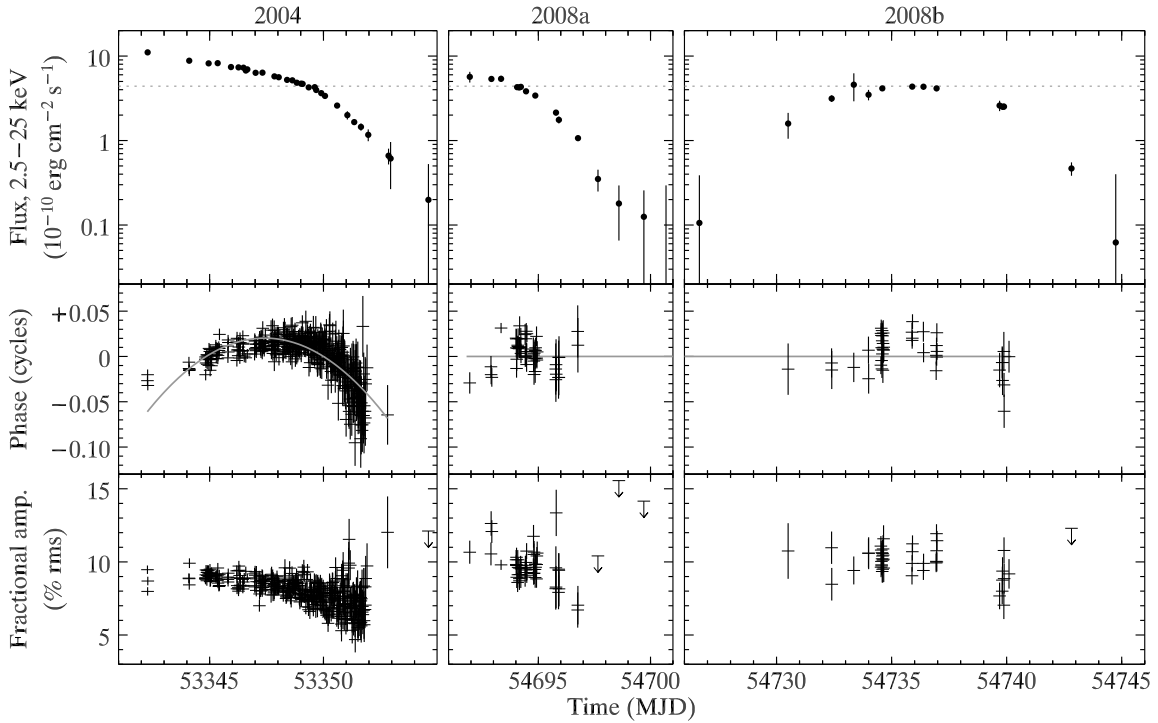}
  \end{center}
  \caption{ The light curves, phases, and fractional amplitudes of the
    observed outbursts of \igrj.  The light curves show the 2.5--25~keV flux,
    with one point per observation, after subtracting the contribution from
    instrumental background and the nearby intermediate polar V709~Cas.  The
    dotted gray line indicates the critical flux $f_{x,c}$, which marks the
    quickening of the decay during the 2004 and 2008a outbursts and the
    maximum during the 2008b outburst.  The pulse phases show the residuals
    measured relative to the best-fit 2008 frequency, $\nu_0 =
    598.892130804$~Hz.  The gray lines show the best-fit models for each
    outburst.  Phases and fractional amplitudes are shown for the 2.5--15~keV
    band.  95\% upper limits on the fractional amplitudes are indicated for
    observations without detectable pulsations; these upper limits for the
    first and last observations of the 2008b outburst were very weak
    ($\sim$50\% rms).
  \label{fig:BigPlot}}
\end{figure*}

For our coherent timing analysis, we included 2--15~keV photons to maximize
the signal to noise ratio.  We shifted the photon arrival times to the solar
system barycenter using the DE405 solar system ephemeris and the optical and
near-infrared position given by \citet{Torres08}: R.A. = $00^{\rm h}29^{\rm
  m}03\fs05\pm0\fs01$, decl. = $+59\degr34\arcmin18\farcs93\pm0\farcs05$
(J2000.0).  This position is consistent with the \Chandra X-ray position
\citep{Paizis05} and the earlier optical position of \citet{Fox04}, which was
used to derive the pulse timing ephemeris in \citet{Galloway05}.\footnote{This
  R.A. differs by $-0\fs032$ (3.2\sig) relative to the radio position of
  \citet{Rupen04}, which was used to derive the pulse timing ephemerides of
  \citet{Falanga05} and \citet{Burderi07}.  If these ephemerides had used the
  optical position, their frequencies would shift by $\Delta\nu =
  4\times10^{-8}$~Hz and $\Delta\dot\nu = 1.0\times10^{-14}$\HzPerSec.  This
  offset in $\dot\nu$ is far too small to account for the spin-ups reported by
  these authors.}  After barycentering the photon arrival times, we applied
the \RXTE fine clock correction and filtered out data during Earth
occultations and intervals of unstable pointing.  We searched the data for
thermonuclear (type I) X-ray bursts; none were present.

To measure the times of arrival (TOAs) and fractional amplitudes of the
persistent pulsations, we folded 626~s intervals of data (twice the V709~Cas
period) using the \citet{Galloway05} timing model.  The timing models were
applied and fitted using the TEMPO pulse timing program, version
11.005,\footnote{At http://wwww.atnf.csiro.au/research/pulsar/tempo} and
assumed a circular orbit and a fixed spin frequency.  From the resulting
folded pulse profiles, we measured the phases and amplitudes of the
fundamental harmonic of the pulsation.  Higher harmonics had insufficient
power to be useful for pulse phase timing.  Estimation of phase and fractional
amplitude uncertainties follow the procedures described in \citet{Hartman08}.

We also analyzed the aperiodic variability during the \igrj outbursts to
characterize its broadband noise properties and search for quasi-periodic
oscillations (QPOs).  We followed the procedures and conventions of
\citet{Linares07}, who measured the timing spectrum of the 2004 outburst, so
that our results would be directly comparable.  We selected 2.5--30~keV
photons, as per the earlier study, and modified their arrival times using our
orbital ephemeris to shift them into the frame of the NS.  We then performed
Fourier transforms on 1024~s segments of data and normalized the resulting
power spectra using the rms normalization of \citet{VanDerKlis95}.  We
estimated the V709~Cas contribution as previously described and corrected for
it and the instrumental background.  We averaged the power spectra to create a
single spectrum for each outburst, then fit them with multiple Lorentzians
following the conventions of \citet{Belloni02}.

\section{Results}

\subsection{Outburst Light Curves and Fluences}
\label{sect:LightCurveResults}

The light curve of the first 2008 outburst showed a fast rise and slow decay.
The rise of the outburst occurred in less than 3.5 days, the length of time
separating the observations during the PCA monitoring campaign.  A monitoring
observation on MJD 54688.4 gave a 2.5--25~keV flux from the field of
$(0.72\pm0.06)\times10^{-10}$\fluxunits, consistent with the flux of V709~Cas.
By the next monitoring observation, on MJD 54691.9, the total 2.5--25~keV flux
had risen to the outburst maximum of $(6.4\pm0.8)\times10^{-10}$\fluxunits,
implying a peak flux from \igrj of $5.7\times10^{-10}$\fluxunits.  The daily
averages from the \Swift Burst Alert Telescope (BAT) suggest that this rise
occured in a single day,\footnote{BAT transient monitoring provided by the
  \Swift/BAT team at
  \url{http://swift.gsfc.nasa.gov/docs/swift/results/transients/}.} although
their large uncertainties ($\sim$$3\times10^{-10}$\fluxunits) limit the
significance of this measurement.  After $\approx$3~d near its maximum, the
first outburst decayed with an $e$-folding time of $(1.8\pm0.3)$~d, returning
to quiescence in $\approx$5~d.  The 2.5--25~keV fluence of the first outburst,
after subtracting a $0.68\times10^{-10}$\fluxunits mean contribution from
V709~Cas, was $(3.0\pm0.1)\times10^{-4}$~erg~cm$^{-2}$.  Most of the
uncertainty in this figure reflects the uncertainty in the V709~Cas flux.

Following the first outburst, \igrj returned to quiescence for 30~d.  The mean
2.5--25~keV flux from the PCA field centered on \igrj was
$0.70\times10^{-10}$\fluxunits during this period, with an rms scatter of
$0.08\times10^{-10}$\fluxunits.  This flux and scatter are consistent with the
monitoring observations before and after the 2008 outbursts and with the
expected V709~Cas contribution.

Unlike the first outburst, the second had an approximately symmetric light
curve.  In $\approx$7~d it rose from quiescence to a peak 2.5--25~keV source
flux of $(4.3\pm0.2)\times10^{-10}$\fluxunits, remained at this peak flux for
2--3~d, then dimmed back to quiescent levels over the subsequent $\approx$7~d.
Observations during the decay were too sparse to determine whether the decay
was exponential.  Despite its qualitatively different light curve, the second
outburst had approximately the same fluence as the first:
$(3.1\pm0.1)\times10^{-4}$~erg~cm$^{-2}$.

A critical flux of $f_{x,c} = 4.4\times10^{-10}$\fluxunits (2.5--25~keV)
played an important role in the three outbursts observed from \igrj.  The top
panels of Figure~\ref{fig:BigPlot} show their light curves, with $f_{x,c}$
marked by the dotted horizontal line.  The 2004 outburst was the brightest,
longest, and most frequently observed by \RXTE.  While its flux was above
$f_{x,c}$, it decayed linearly at a rate of
$(-0.89\pm0.01)\times10^{-10}$\fluxunits~d$^{-1}$.  Upon reaching $f_{x,c}$,
the decay quickened and became exponential, with an $e$-folding time of
$(1.9\pm0.2)$~d (in reasonable agreement with the 2.2~d time scale reported by
\citealt{Falanga05} in their analysis of the \INTEGRAL light curve of this
outburst).  The critical flux played a similar role in the 2008a outburst:
the slow decay seen during the first 3~d of the outburst quickened
considerably after crossing below $f_{x,c}$.  Finally, the symmetric light
curve of the 2008b outburst slowly rose to a maximum flux of
$\approx$$f_{x,c}$, then reversed and slowly decayed.  The implications of
this critical flux will be considered in Section~\ref{sect:ModelingDoubleOB}.

\begin{figure}[t!]
  \begin{center}
    \includegraphics[width=0.45\textwidth]{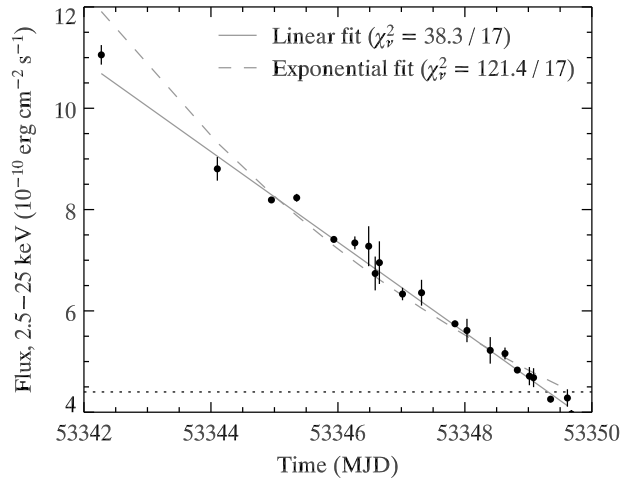}
  \end{center}
  \caption{ Light curve for the first 7~d of the 2004 outburst and a
    comparison the best-fit linear (solid line) and exponential (dashed line)
    models.  The dotted line at bottom indicates $f_{x,c}$.
  \label{fig:LinearVsExpDecay}}
  \vspace{1em}
\end{figure}

The linear decay that we report during the first 7~d of the 2004 outburst is
contrary to theoretical expectations, and it therefore merits close
examination.  The accretion disk model of \citet{Powell07}, which we discuss
in detail in Section~\ref{sect:HotDiskModels}, predicts an exponential decay
for this part of the outburst, and \citet{Falanga05} reported an exponential
fit to these data.  To parameterize the shape of this decay, we use the
following model:
\begin{equation}
  f_x = \left(f_{x,0}^{-\xi} +
    f_{x,c}^{-\xi} \cdot \xi \frac{t}{\tau}\right)^{-1/\xi} \, .
\end{equation}
We fit the initial flux $f_{x,0}$ at time $t = 0$, the decay time scale
$\tau$, and the shape parameter $\xi$.  This choice of model will be motivated
by our investation of irradiated accretion disks in
Section~\ref{sect:HotDiskModels}, but for now note that it encompasses linear
($\xi = -1$), exponential (the limit $\xi \to 0$), and $1/t$ decay ($\xi =
1$).

The brightest portion of the 2004 light curve is consistent with linear decay:
$\xi = -1.05\pm0.13$ with $\chi^2 = 38.2$ for 16 degrees of freedom.  In
comparison, forcing an exponential model (i.e., holding $\xi$ fixed at very
nearly zero) gives $\chi^2 = 121.4$ for 17 degrees of freedom.
Figure~\ref{fig:LinearVsExpDecay} compares these two fits.  Some irradiated
accretion disk models have $\xi = 1/4$, which fares even worse: $\chi^2 =
177.0$ for 17 degrees of freedom.  The time scale for the linear fit is $\tau
= 4.9\pm0.1$~d, which is the length of time needed for the flux to fall by the
amount $f_{x,c}$.  This $\tau$ is similar to the 6.6~d $e$-folding time scale
reported by \citet{Falanga05} for an exponential fit.  Note that we can
exclude an exponential decay even without the outlying first observation (on
MJD 53342), since the $\chi^2$ for an exponential fit remains roughly three
times the linear $\chi^2$ without this point.

\subsection{Spectral Evolution during the Outbursts}
\label{sect:SpecParams}

The unusual light curves of the 2008 outbursts of \igrj naturally raise the
question of whether its spectrum undergoes significant change during the
outbursts.  In particular, does the spectrum show any transition accompanying
the change in the decay rate at $f_{x,c}$?

Other sources with ``knees'' in their light curves do not have associated
abrupt spectral changes.  \RXTE observations of AMSPs are generally well fit
by an absorbed $\sim$1~keV blackbody plus power law.  In the case of \saxj,
the power law index gradually softened from 1.5 to 1.9 over the courses of the
outbursts \citep{Gierlinski02, Ibragimov09}, but no significant change in this
trend coincided with that source's transition from slow to rapid decay
\citep{Hartman09a}.  For SWIFT~J1756.9$-$2508, the photon index softened from
1.8 to 2.0 \citep{Linares08}.  XTE~J0929$-$314 and XTE~J1751$-$305 had roughly
constant spectral parameters, with no change across the knees of their light
curves \citep{Juett03, Gierlinski05}.  Analysis of the 2004 outburst of \igrj
by \citet{Paizis05} reported an unchanging spectrum: their fits used a
$\sim$1~keV blackbody component and a constant photon index of 1.6.  They also
found hints of a 6.4~keV iron line.

To measure the spectrum of \igrj, we first fit the contribution from the
nearby intermediate polar V709~Cas.  For the 2004 outburst, we used the \RXTE
observations of the \igrj field following its quiescence, and for 2008 we used
the observations during the 30~d quiescence between outbursts.  In both cases
we fit V709~Cas with an absorbed thermal bremsstrahlung spectrum plus a narrow
iron line fixed at 6.4~keV.  The resulting plasma temperatures of $26\pm5$~keV
in 2004 and $31\pm4$~keV in 2008 agree with earlier findings
\citep{DeMartino01, Falanga05b}.  When performing subsequent fits of \igrj, we
held the parameters of the V709~Cas contribution fixed for each epoch,
effectively subtracting its contribution from the total flux.  To reduce the
uncertainty introduced by potential errors in our V709~Cas fits, we only
measured the spectral parameters of \igrj using observations in which it was
brighter than V709~Cas.

\begin{figure}[t!]
  \begin{center}  
    \includegraphics[width=\columnwidth]{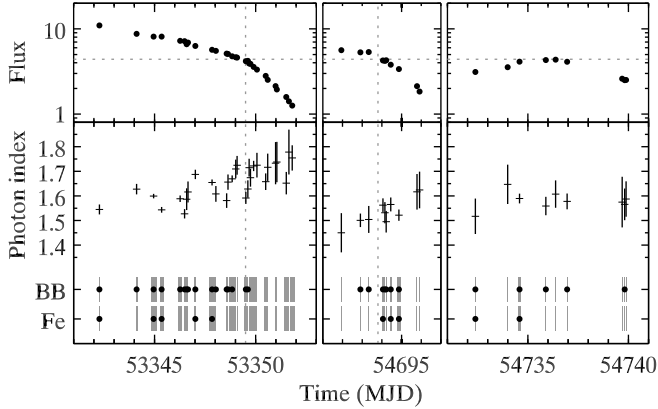}
  \end{center}
  \caption{ The spectral fits for the three outbursts.  The 2.5--25~keV flux
    from \igrj is shown for reference, in units of
    $10^{-10}$~erg~cm$^{-2}$~s$^{-1}$.  Dotted gray lines mark the critical
    flux $f_{x,c}$ and the times at which the outbursts cross this flux.  The
    lower plot shows the fit components: a predominant power law and an
    intermittently detected $\sim$1~keV blackbody and 6.4~keV iron line.  The
    blackbody and iron line components are marked with circles where they are
    detected with at least 90\% confidence; accompanying vertical bars show
    the times of observations, with the width of the bars showing their
    lengths.
  \label{fig:SpecParams}}
\end{figure}

We fit the \igrj flux using an absorbed blackbody plus power law, with the
addition of a 6.4~keV iron line when needed.  The absorption column was held
fixed at $n_H = 4.6\times10^{21}$~cm$^{-2}$ based on the optical work of
\citet{Torres08}.  Figure~\ref{fig:SpecParams} shows the key results.  The
blackbody temperature was consistent with a constant 0.9~keV.  In agreement
with the analysis of \citet{Paizis05}, we found that this blackbody component
fades more rapidly than the power law component during 2004, becoming
insignificant around the time when the total flux falls below $f_{x,c}$;
however, this is not the case for the 2008 outbursts, for which the ratio of
the blackbody and power law contributions remains roughly constant throughout.
We found that the power law component softened during the 2004 and 2008a
outbursts.  Data for the 2008b outburst are limited but are consistent with a
constant power law index.  This softening during 2004 is contrary to the
results reported in \citet{Paizis05}.  This discrepancy was likely due to an
insufficient compensation for the harder spectrum of V709~Cas: if its
contribution is not subtracted, the overall power law index is indeed roughly
constant.  These 2008 spectra are generally compatible with the two
contemporaneous Swift XRT detections, which yielded power law indices of
$\approx$1.6 and did not require blackbody components \citep{Lewis10}.

No abrupt changes in the spectrum coincide with the break in the light curve
decay at $f_{x,c}$.  The rate at which the photon index gradually increases
does not change at this point, and no other spectral parameter is
significantly affected by this transition.

\subsection{Pulse Timing during the 2008 Outbursts}

The pulse TOAs during the 2008 outbursts exhibit a low level of timing noise
relative to other AMSPs.  The rms amplitude of the phase residuals of the
best-fit constant-frequency model is 0.016~rotational cycles (27~\us).
Counting noise accounts for approximately half of these residuals,
contributing an rms amplitude of 0.011~cycles.  The remaining intrinsic timing
noise does not exhibit any long-period trends or abrupt phase shifts (as in
\saxj; \citealt{Burderi06, Hartman08}) or correlations with the flux (as in
XTE~J1814$-$338; \citealt{Papitto07, Patruno09c}).  This greatly simplifies
analysis, allowing us to estimate the uncertainties of timing model parameters
by scaling the phase uncertainties such that the reduced $\chi^2$ of the model
is unity.  (Because the initial phase uncertainties assume only counting
noise, which accounts for roughly half the total noise, the scaling factors
were consistently $\approx$$\sqrt{2}$.)  A separate bootstrapping analysis
supports the validity of this approach by producing similar model
uncertainties.

We first developed separate timing models for each of the 2008 outbursts.
These models and the amplitudes of their phase residuals are summarized in
Table~\ref{tbl:FreqModels}, with a constant-$\nu$ model and a non-zero
$\dot\nu$ model fit for each outburst.  Including a frequency derivative
yields a spin-up for both outbursts, but at low significance: 1.0\sig for
2008a and 1.8\sig for 2008b.  The large uncertainties on these frequency
derivatives are due to the sparsity of observations, particularly during the
beginnings and ends of the outbursts when phase measurements have the greatest
leverage for fitting $\dot\nu$, and the decreased effective area of the PCA.
While the 2008 data are adequately fit by a constant-$\nu$ model, we cannot
conclusively rule out the presence of a spin-up of the magnitude observed
during the 2004 outburst, $8.4(6)\times10^{-13}$\HzPerSec \citep{Falanga05}.

The proximity of the 2008 outbursts enables phase connection between two AMSP
outbursts for the first time.  With frequency accuracies of $\sim$30~nHz for
the two outbursts, the phase uncertainty when connecting across the 30~d
interim of quiescence is $\sim$0.05~cycles --- small enough to ensure that we
do not miss or include an extra rotation of the neutron star.  This long
baseline greatly improves the accuracy of our pulse timing model, and the best
constant-frequency fit is given in Table~\ref{tbl:FreqModels}.  This frequency
is consistent with the frequencies measured for the individual outbursts.  The
addition of a frequency derivative does not significantly improve the fit.

\begin{deluxetable}{lccc}
\tabletypesize{\footnotesize}
\tablecolumns{2}
\tablewidth{0pt}
\tablecaption{Frequency models for the 2008 outbursts
  \label{tbl:FreqModels}}
\tablehead{
  \colhead{} &
  \colhead{$\nu$} &
  \colhead{$\dot\nu$} &
  \colhead{Residuals} \\
  \colhead{} &
  \colhead{(Hz)} &
  \colhead{($10^{-13}$\HzPerSec)} &
  \colhead{(rms cycles)}
}
\startdata
2008a  &    598.89213081(4)   &    ---     &  0.0148\\
       &    598.89213081(4)   &   10(10)   &  0.0133\\[0.5em]
2008b  &    598.89213084(3)   &    ---     &  0.0175\\
       &    598.89213084(3)   &  4.5(2.5)  &  0.0150\\[0.5em]
Both   & \pz598.892130804(2)  &    ---     &  0.0163\\[-0.75em]
\enddata
\tablecomments{For models with a non-zero $\dot\nu$, $\nu$ is specified in the
  middle of the outburst.\\}
\end{deluxetable}

This phase connection between outbursts does require the assumption that
torques during the 30~d quiescence are small.  A quiescent spin derivative of
$|\dot\nu| \gtrsim 3\times10^{-13}$\HzPerSec would introduce an unknown number
of phase wraps.  We will address whether such torques are physical in the
discussion.  It is worth noting that the phases of the two 2008 outbursts line
up quite well.  If we extrapolate the phase solution of the 2008a
constant-frequency model forward and apply it to the 2008b outburst without
any additional fitting, the residuals during 2008b have a mean offset of only
+0.02~cycles relative to our solution.  If phase wraps occurred between 2008a
and 2008b, then this excellent alignment would be purely coincidental.

Finally, we consider the possibility that the marginal spin-ups during the
2008 outbursts are real.  For concision, we compare their effects relative to
the best-fit constant frequency during 2008, $\nu_0 = 598.892130804$~Hz.  The
values of $\dot{\nu}$ listed in Table~\ref{tbl:FreqModels} would result in a
frequency of $\nu = \nu_0 + 0.23(13)$~\uHz at the end of 2008a (at MJD
54697.0) and $\nu = \nu_0 - 0.24(11)$~\uHz at the beginning of 2008b (at MJD
54730.0).  The neutron star would have to undergo a mean spin-down of
$\dot{\nu} = -1.7(6)\times10^{-13}$\HzPerSec during the intervening 33 days of
quiescence to account for this decrease of frequency.  If a spin-up is present
during these outbursts, then a spin-down of a similar magnitude necessarily
follows them.

\subsection{Long-term Spin and Orbital Evolution}

Despite the highly accurate (2~nHz uncertainty) timing solution attained by
phase connecting the 2008 outbursts, it is still not feasible to phase connect
the 2004 and 2008a outbursts.  During the 3.6~years of quiescence separating
these outbursts, magnetic dipole torques from the neutron star's unknown field
strength (say, $10^{7.5}$--$10^{8.5}$~G) will produce 1--50 phase wraps
relative to a model with a constant frequency.  This prevents the construction
of a coherent timing model for all the data.

Comparing the frequencies of the outbursts is more useful.  From the ephemeris
of \citet{Falanga05}, the frequency at the end of the 2004 outburst will be
$\nu = \nu_0 + 0.32(4)$~\uHz; the model of \citet{Burderi07} and our own
analysis give similar figures.  As discussed in the previous section, the
determination of $\dot\nu$ for the 2008a outburst is more ambiguous.  If we
accept the marginally significant spin-up, then the frequency at the beginning
of the 2008a outburst is $\nu = \nu_0 - 0.21(19)$~\uHz; if not, then we can
phase connect the 2008 outbursts to give the constant frequency of $\nu_0$
with an uncertainty of 0.002~\uHz.  These scenarios produce respective
frequency drops of 0.53(20)~\uHz and 0.32(4)~\uHz during the 3.6~years of
quiescence.  Averaged over the entire quiescent period, these drops represent
a mean spin-down of $-(3$--$5)\times10^{-15}$\HzPerSec.

Connecting the orbital phases of the 2004 and 2008 outbursts provides a great
increase in the accuracy of the orbital period.  Table~\ref{tbl:Orbit}
summarizes the orbital ephemeris derived by combining the data from all three
outbursts.  Detecting an orbital period derivative is possible in principle
with three orbital phase measurements, but the proximity of the 2008 outbursts
prevents anything more than a weak upper limit on the orbital period
derivative.  The effect of any physically plausible derivative on the orbital
phase during the 30~d separating the outbursts would be too small to measure.

\begin{deluxetable}{@{} l @{~} r @{}}
\tabletypesize{\footnotesize}
\tablecolumns{2}
\tablewidth{0pt}
\tablecaption{Orbital parameters for \igrj \label{tbl:Orbit}}
\startdata
\hline\hline\\[-1.5 ex]  
Orbital period, $P_{\rm orb}$ (s) &
  8844.076729(9)\\
Orbital period derivative, $\dot{P}_{\rm orb}$ (95\% UL; s~s$^{-1}$) &
  $< 3\times10^{-11}$\\
Projected semimajor axis, $a_{\rm x} \sin i$ (light-ms) &
  64.9911(4)\\
Time of ascending node, $T_{\rm asc}$ (MJD, TDB) &
  54000.0739971(7)\\
Eccentricity, $e$ (95\% confidence upper limit) &
  $< 2\times10^{-4}$\\[-0.75em]
\enddata
\end{deluxetable}

\subsection{Pulse Profiles}

The pulses from \igrj were entirely consistent with a sinusoid across the
2--60~keV energy band of the PCA.  A relatively low amount of timing noise
allowed us to integrate the pulse profile over both 2008 outbursts.  Folding
all the 2.5--15~keV photons gave a profile that was well-fit by a pure
sinusoid with no harmonics ($\chi^2 = 259.8$, 253~degrees of freedom).  The
fractional amplitude in this band was 10\%~rms and remained roughly constant
throughout both outbursts.  The 95\% upper limits on the fractional amplitude
for the second and third harmonics were both 0.36\%~rms, or $\lesssim$4\% of
the fundamental's amplitude.  These results are consistent with the
non-detection of harmonics during the 2004 outburst \citep{Galloway05,
  Falanga05}.

\begin{figure}[t!]
  \begin{center}
    \includegraphics[width=0.45\textwidth]{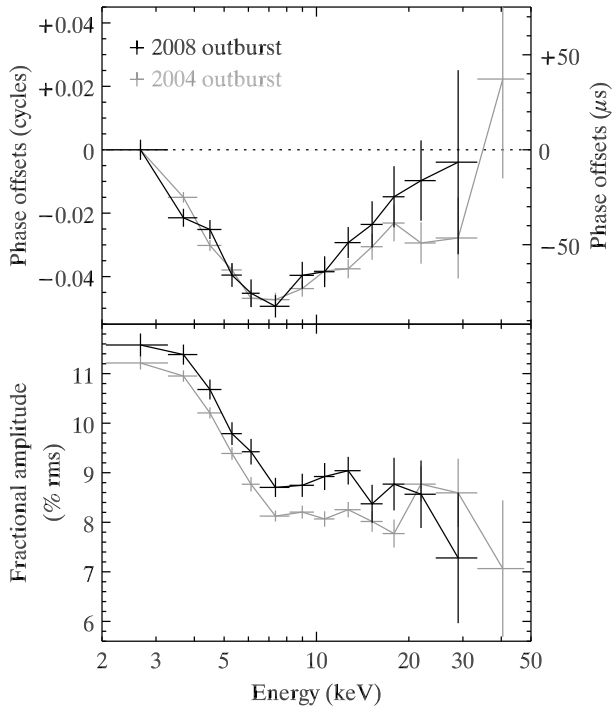}
  \end{center}
  \caption{ The phases and fractional amplitudes as a function of energy.  The
    black points are integrated over both 2008 outbursts; the gray points are
    for the 2004 outburst.  Error bars show the statistical uncertainties.
    For the fractional amplitudes, an additional uncertainty in the overall
    vertical offset is introduced by the approximated flux from V709~Cas.
  \label{fig:EnergyDep}}
\end{figure}

To measure the energy dependence of the pulsations, we divided the PCA
response into 15 energy bands and folded all the 2008 outburst data for each.
Figure~\ref{fig:EnergyDep} shows the measured pulse arrival times and
fractional amplitudes.  Data from the 2004 outburst are also shown for
comparison.  Pulsations were detected with 99\% confidence at 2--35~keV
(2--48~keV during the 2004 outburst).

From 2--8~keV, the pulse arrival times follow the usual pattern of soft lags
that is seen in most other AMSPs.  As \citet{Falanga07} noted for the 2004
outburst, above 8~keV the trend reverses and harder bands increasingly lag.
Agreement between the 2004 and 2008 phases is generally good.  There was no
significant change in the magnitude of the lags over the course of the
outbursts, in contrast to the flux dependence of the lags observed in \saxj
\citep{Hartman09a}.

The fractional amplitude decreases significantly until 8~keV, above which the
energy dependence is weaker but with some evidence of further decrease.  The
error bars for both the phases and amplitudes only account for statistical
uncertainty.  The fractional amplitude errors do not include the uncertainty
in the flux from V709~Cas, which is assumed to be the same for both outbursts.
Nevertheless, the offset between the 2004 and 2008 amplitudes is significant.
The V709~Cas flux would have to be 30\% higher during 2004 to account for this
fractional amplitude difference, but a comparison of its flux from monitoring
observations after the 2004 and 2008 outbursts returned to quiescence show
that this was not the case.

\begin{figure}[t]
  \begin{center}
    \includegraphics[width=0.45\textwidth]{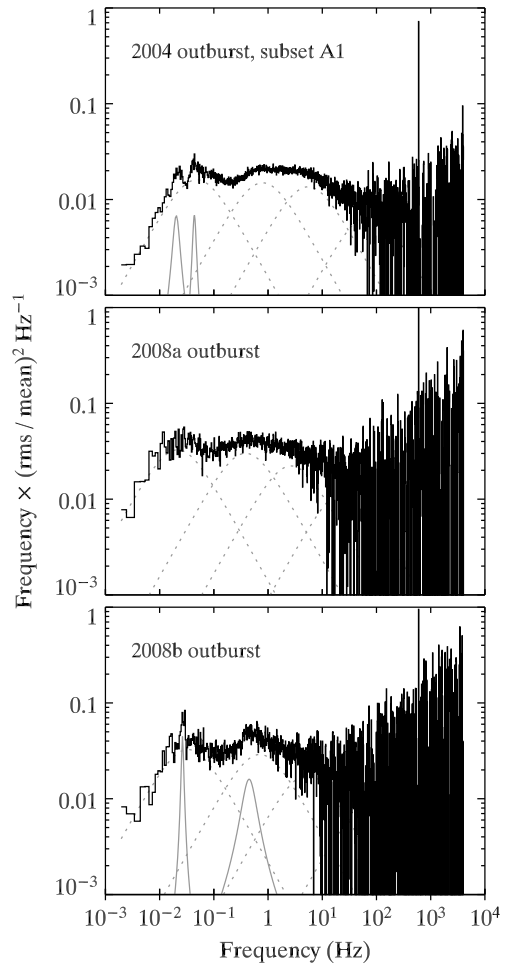}
  \end{center}
  \caption{ The timing spectra of the three outbursts from \igrj.  Dotted gray
    curves show Lorentzian components centered at the origin (i.e., with a
    coherence of $Q = 0$); solid gray curves show QPO components, modeled with
    Lorentzians with non-zero coherences.  Subset A1 of the 2004 outburst
    includes only the brightest part of that outburst, as defined by
    \citet{Linares07}.
  \label{fig:PowerSpecs}}
\end{figure}

No pulsations were detected during the quiescent period between the 2008
outbursts.  Folding all data in MJD 54700--54730 yields a 95\% upper limit on
the pulse fractional amplitude of 0.4\%~rms, accounting for instrumental
background only.  This limit further supports the assumption that the
non-background photons during this period are from V709~Cas, and \igrj is
indeed in quiescence.

\begin{deluxetable*}{lccc}
\tabletypesize{\footnotesize}
\tablecolumns{4}
\tablewidth{0pt}
\tablecaption{Aperiodic timing properties of \igrj \label{tbl:AperTiming}}
\tablehead{
  \colhead{} &
  \colhead{2004\tablenotemark{ a}} &
  \colhead{2008a} &
  \colhead{2008b}
}
\startdata
Interval of spectrum (MJD)  &  53341.0--53348.2  &  54694.0--54695.0  &  54733.0--54737.5 \\
Min/mean/max flux ($10^{-10}$\fluxunits) & 5.4 / 6.8 / 10.9 & 4.0 / 4.2 / 4.8 & 3.5 / 4.2 / 4.8 \\
Total 0.01--100~Hz variability (\% rms) & $38.8\pm0.2$ & $50.5\pm0.5$ & $51.3\pm0.7$ \\
Low-frequency break ($10^{-2}$~Hz)\tablenotemark{b} & $4.9\pm0.2$ & $2.1\pm0.2$ & $3.3\pm0.2$ \\
Lower QPO ($10^{-2}$~Hz) & $2.03\pm0.06$ & --- & $2.64\pm0.05$ \\
Upper QPO ($10^{-2}$~Hz)\tablenotemark{c} & $4.37\pm0.08$ & --- & $45\pm2$\pz\\[-0.75em]
\enddata
\tablenotetext{a}{Subset A1 only.  The 2004 figures are from our fit but agree
  with \citet{Linares07}.}
\tablenotetext{b}{The break frequency given is $\nu_{\rm max}$ (see
  \citealt{Belloni02}) of the lowest-frequency zero-centered Lorentzian.}
\tablenotetext{c}{Unlike the 2004 QPOs, the QPOs during 2008b are not
  related.}
\end{deluxetable*}

\subsection{Aperiodic Timing}
\label{sect:AperiodicTiming}

A broadband power spectrum of \igrj reveals a number of notable features.
Figure~\ref{fig:PowerSpecs} shows the spectra, and Table~\ref{tbl:AperTiming}
lists some of their properties.  For comparison, the top plot of
Figure~\ref{fig:PowerSpecs} shows the power spectrum of the first 6~d of the
2004 outburst, a selection labeled ``subset A1'' in the analysis of
\citet{Linares07}.  High levels of aperiodic timing noise distinguish the
spectra of this source from those of other NS LMXBs.  The noise level is flat
down to very low frequencies ($\sim$0.01~Hz), resulting in integrated
fractional variabilities over 0.01--100~Hz of 40--60\% rms.  Additionally, two
harmonically related quasi-periodic oscillations (QPOs) are present at around
0.02~Hz and 0.04~Hz \citep{Linares07}.

The broadband shape and integrated variability of the 2008 power spectra are
similar.  The spectra are roughly flat down to a break at $\sim$0.01~Hz, and
the integrated fractional variability is $\approx$50\% rms.  No QPOs were
detected during the 2008a outburst, but its higher overall noise level would
be sufficient to bury the QPOs observed during 2004 if they were present at
the same fractional amplitude and coherency ($\approx$5\% rms; $Q \sim 5$), so
their non-detection is not constraining.  During the 2008b outburst, a single
QPO was definitively seen near the low-frequency break.  It is unclear which
harmonic it represents if it is a member of a harmonically related QPO pair:
at 0.026~Hz, it falls between the QPO frequencies seen in the earlier
outburst.  An additional QPO at 0.45~Hz is needed to model the steeper power
spectrum ``hump'' present during 2008b.  This QPO has an amplitude of
$(13\pm2)\%$, a coherency of $Q = 1.5\pm0.5$, and is present in all 2008b
observations with sufficient integration time to detect it.  The 2008 data
support some of the overall trends identified by \citet{Linares07},
particularly the anticorrelation between the overall variability and the flux
and the positive correlation between the break frequency and the flux.

\section{Modeling the Outburst Light Curves}
\label{sect:ModelingDoubleOB}

The most unusual feature of the 2008 activity of \igrj is its light curve:
the source undergoes two outbursts in rapid succession.  We are unaware of any
other NS LMXB for which this is the case.\footnote{Some NS LMXBs do show much
  shorter ($\lesssim3$~d) and dimmer mini-outbursts or flares that recur
  frequently; XTE~J1751$-$305 \citep{Markwardt07b} and NGC~6440~X-2
  \citep{Heinke10} are notable examples.  At the other extreme, the
  intermittently pulsating AMSP HETE~J1900.1$-$2455 has been in outburst since
  2005, with the exception of 1--6~d in 2007 during which it was briefly
  quiescent \citep{Degenaar07}.  Unlike these systems, all the outbursts from
  \igrj resemble the typical $\sim$10~d outbursts seen in most NS LMXB
  transients.}  This double outburst provides a rigorous test of the accretion
disk models put forward to explain the transient nature of many LMXBs.  A
successful theory must accomplish the following:

{\em Explain the light curves of the 2004 and 2008a outbursts, particularly
  the knee at the critical flux $f_{x,c}$.}  These two outbursts were preceded
by a long ($\sim$3~yr) period of quiescence, and they followed a fast rise /
slow decay profile.  Both showed a knee in the decay at a critical 2.5--25~keV
flux of $f_{x,c} = 4.4\times10^{-10}$\fluxunits.  During the 2004 outburst,
the decay prior to the knee was nearly linear, with a timescale of $f_{x,c} /
\dot{f}_x = 4.9$~d; during the 2008a outburst, a lower peak flux and sparser
sampling prevented characterization of the nature of the pre-knee decay.  Both
outbursts showed exponential decays after the knee with $e$-folding times of
$\approx$1.9~d.

{\em Stop the 2008a outburst before the disk becomes too depleted to fuel the
  2008b outburst.}  Unless the mass transfer rate from the companion can vary
by orders of magnitude on a $\sim$1~month time scale, the mass needed for
2008b could not have accumulated during the 30~d quiescence that preceded it.
The 2.5--25~keV fluence of the 2004 outburst was
$5.2\times10^{-4}$~erg~cm$^{-2}$~s$^{-1}$, and the summed fluence of the 2008
outbursts was similar, at $6.1\times10^{-4}$~erg~cm$^{-2}$~s$^{-1}$.  Taking
these fluences and a 3~yr recurrence time as typical, we estimate that at
least 95\% of the matter accreted during 2008b must have been present in the
disk at the end of 2008a.

{\em Account for the symmetric light curve of 2008b, and why its peak flux is
  $f_{x,c}$.}  In contrast to outbursts following a long period of quiescence,
the 2008b outburst rose slowly over $\approx$7~d until reaching a maximum flux
of $f_{x,c}$.  After 2--3~d at this maximum flux, the outburst then slowly
decayed back to quiescent levels over another $\approx$7~d.

We consider this unique phenomenology in light of three models: the disk
diffusion model of \citet{Wood01}, which was developed to explain a similar
double outburst from the black hole transient XTE~J1118$+$480; accretion from
a partially ionized irradiated disk; and accretion at rates near the onset of
a quasi-propeller state, in which the centrifugal acceleration of infalling
matter by the NS magnetosphere inhibits but does not halt accretion.

\subsection{Disk Diffusion Model}

The one other LMXB with a published report of a double outburst is the black
hole transient XTE~J1118$+$480.  Its first outburst lasted 40~d and followed a
fast rise / exponential decay profile.  After 30~d quiescence, a second
outburst lasted 150~d and had an irregular profile with multiple peaks
\citep{Wood01}.

Additionally, \igrj and XTE~J1118$+$480 have similar aperiodic timing
properties.  \citet{Linares07} note that the strong very-low-frequency
($\lesssim 0.1$~Hz) variability and high overall variability of \igrj much
more closely resemble the timing spectra of black hole LMXBs than other NS
LMXBs.  In section~\ref{sect:AperiodicTiming}, we reported a 0.45~Hz QPO
during the 2008b outburst of \igrj; scaled for the difference in central
object masses (1.4~$M_\sun$ vs. 6--7~$M_\sun$), it matches the $\sim$0.1~Hz
QPO observed during the second outburst of XTE~J1118$+$480 \citep{Wood00}.
The data for the 2008b outburst were not adequate to determine whether the QPO
frequency increased over the course of the outburst, as was observed for
XTE~J1118$+$480.

Given the observational similarities between these sources, it is worth
considering the disk diffusion model of \citet{Wood01}.  Noting that the disk
instability model was insufficient to explain the double outburst of
XTE~J1118$+$480, they proposed that the light curve followed from a varying
mass loss rate from the companion, which would accrete onto the NS after
diffusing through the accretion disk.  The resulting light curve would be the
convolution of the companion's mass loss function and a fast-rise,
exponential-decay diffusion function that acts on the viscous time scale of
the disk.  It is unlikely that this model can account for the outbursts of
\igrj, however.  If the viscous time scale of the disk is similar to the decay
time scales of the 2004 and 2008a outbursts, then their observed light curves
would require a very rapid ($\lesssim$3~d) transfer from the companion to the
outer accretion disk of most of the mass consumed in those outbursts.
Furthermore, a disk diffusion model cannot explain the knee in the light
curves.  The 2008b outburst would require only slightly slower transfer of
matter from the companion to the disk.

\subsection{Accretion from an Irradiated Disk}
\label{sect:HotDiskModels}

\newcommand{\disk}{\mathrm{disk}}
\newcommand{\hot}{\mathrm{hot}}
\newcommand{\cold}{\mathrm{cold}}
\newcommand{\visc}{\mathrm{visc}}
\newcommand{\co}{\mathrm{co}}
\newcommand{\bol}{\mathrm{bol}}

We next consider the outbursts from the perspective of the disk instability
model as applied to soft X-ray transients by \citet[][hereafter KR98]{King98}
and \citet[][hereafter P07]{Powell07}.  In this approach, the accretion disk
is divided into two regions: a hot inner disk in which ionization results in a
high viscosity and the inward migration of matter; and a cooler, non-ionized
outer disk in which the inward drift is much lower.

The radial extent of the hot, ionized region can be calculated by considering
the temperature of the disk due to irradiation from the NS.  If an accretion
disk with a scale height $H \propto R^n$ and albedo $\eta_*$ is irradiated by
a small central source with a luminosity $L_x$, \citet{DeJong96} calculate its
temperature profile to be
\begin{equation}
  T^4 = \frac{1 - \eta_*}{4\pi R^2} \frac{H}{R} (n-1) L_x \, .
  \label{eq:DiskTemp}
\end{equation}
Following the argument of P07, the hot region will extend out to some
temperature $T_h$ at which the gas in the disk becomes mostly ionized.  If we
denote the radius of this temperature as $R_h$, we can solve to find
\begin{equation}
  R_h^{3-n} \propto L_x \propto \dot{M}_x \, ,
\end{equation}
where $\dot{M}_x$ is the accretion rate onto the compact object.  When the
disk is illuminated by a point-like central source, as is the case here, $n
\approx 9/7$ (KR98).

Above some critical accretion rate $\dot{M}_{x,c}$, the luminosity will be
sufficient to ionize the entire disk.  Thus we can write
\begin{equation}
  \frac{R_h}{R_\disk} =
    \left(\frac{\dot{M}_x}{\dot{M}_{x,c}}\right)^{1 / (3-n)}
    \!\equiv\, \dot{m}_x^\gamma \, .
\end{equation}
To simplify, we define the dimensionless accretion rate $\dot{m}_x = \dot{M}_x
/ \dot{M}_{x,c}$ and the disk opening-angle parameter $\gamma = 1/(3-n)$.  As
$\dot{M}_x$ falls below $\dot{M}_{x,c}$, the transition from a fully to a
partially ionized disk will generally cause the decay rate to increase,
causing a knee in the light curve.  This mechanism has been invoked to explain
the knee seen in the outburst light curves of many transient LMXBs (e.g.,
P07).

\subsubsection{Assumptions of the Viscosity Model}

The models of KR98 and P07 make two assumptions about the disk viscosity.  (1)
The viscous time scale of the ionized region of the accretion disk is shorter
than the decay time scale of the outburst:
\begin{equation}
  \tau_\visc \sim \frac{R_h^2}{\bar{\nu}} \ \ \lesssim\ \ 
  \left|\frac{d\ln \dot{M}_x}{dt}\right|^{-1}
\end{equation}
Here $\bar{\nu}$ is some mean kinematic viscosity of the ionized region.
(Throughout this discussion we adopt the convention that $\nu$ represents the
kinematic viscosity rather than the NS spin frequency, unless otherwise
noted.)  If this condition is met, the surface density of the hot disk will
relax into the quasi-steady profile
\begin{equation}
  \Sigma_\hot(R) \approx \frac{\dot{M}_x}{3\pi\nu} \, .
  \label{eq:HotDiskProfile}
\end{equation}
(2) The viscous time scale of the cold region is far longer than the decay
time scale, so its surface density profile can be treated as independent of
$\dot{M}_x$.

For simplicity, the calculations of these papers also make the approximation
$n = 1$ (or equivalently $\gamma = 1/2$) for a NS LMXB, and they use a
constant mean kinematic viscosity for $\nu$.  We adopt a more general approach
by keeping $\gamma$ a free parameter and by allowing the disk viscosity to
vary with the radius and accretion rate:
\begin{equation}
  \nu(R, \dot{M}_x) = \nu_0 \, (R / R_\disk)^\beta \, \dot{m}_x^\xi \, .
  \label{eq:ViscParams}
\end{equation}
For the \citet{Shakura73} disk solution, $\xi = 3/10$ and $\beta = -3/4$.  For
an irradiation-dominated disk, we assume $\nu \propto T$, and \eq{DiskTemp}
gives $\xi = 1/4$ and $\beta = (n-3)/4 \approx -3/7$.

\subsubsection{Light Curve Decay of a Fully Ionized Disk}
\label{sect:IonizedDiskDecay}

Following the arguments of KR98, we can analytically solve for the light curve
when the accretion disk is entirely ionized:
\begin{equation}
  \dot{M}_x = \left(\dot{M}_{x,0}^{-\xi} + 
    \dot{M}_{x,c}^{-\xi} \cdot \xi\frac{t}{\tau_\hot}\right)^{-1/\xi} \, .
  \label{eq:HotDiskDecay}
\end{equation}
The initial accretion rate is $\dot{M}_{x,0}$ at time $t = 0$, and the time
scale for decay is
\begin{equation}
  \tau_\hot = \frac{2(1 - \xi)}{3(2 - \beta)} \frac{R_\disk^2}{\nu_0} \, .
  \label{eq:TauHotDef}
\end{equation}
Note that in the limit of a uniform and constant viscosity ($\beta = \xi =
0$), these equations reduce to the exponential decay given in KR98: $\dot{M}_x
= \dot{M}_{x,0}\, \exp(-t \cdot 3\nu/R_\disk^2)$.

In this model, a knee in the decay of the outburst is generally attributed to
a transition from a fully to partially ionized disk.  However, it is not
possible to reconcile this hypothesis with the linear decay observed when $f_x
> f_{x,c}$ during the 2004 outburst of \igrj.  Fitting \eq{HotDiskDecay} to
the light curve, we find $\xi = -1.05\pm0.13$, in strong disagreement with the
expected value of $\xi = 0.25$ for an irradiated disk.  Therefore the
accretion disk of \igrj is not fully ionized during the initial decay of the
2004 and 2008a outbursts, and the significance of $f_{x,c}$ must be otherwise
explained.

\subsubsection{Behavior of a Partially Ionized Disk}

When the disk is only partially ionized, we consider the mass budget of the
hot disk, following P07:
\begin{equation}
  \dot{M}_\hot =
    -\dot{M}_x + \mu_\cold(R_h) + 2\pi R_h \dot{R}_h \Sigma(R_h) \, .
  \label{eq:HotDiskBudget}
\end{equation}
The $\dot{M}_\hot$ term can be derived by integrating the hot-disk surface
density out to $R_h$, then taking the time derivative.  From assumption 1,
this surface density follows \eq{HotDiskProfile}.  The $\mu_\cold(R_h)$ term
gives the viscously driven inward flow of matter from the cold disk at $R_h$.
From assumption 2, the radial flow in the cold disk is far smaller than in the
hot disk, so we can neglect this term.  The $\Sigma(R_h)$ term reflects the
addition of cold-disk matter due to the encroaching ionization radius during
the outburst rise, and conversely the return of matter to the cold disk as
$R_h$ recedes during the outburst decay.

Applying these relations to \eq{HotDiskBudget} and casting it in terms of the
dimensionless accretion rate $\dot{m}_x = \dot{M}_x / \dot{M}_{x,c}$ gives a
general model for the partially ionized disk:
\begin{equation}
  \ddot{m}_x = \left[ 2\pi \gamma \dot{m}_x^{2\gamma-2}
    \frac{R_\disk^2 \Sigma(R_h)}{\dot{M}_{x,c}}
    - \frac{1-\zeta}{1-\xi} \tau_\hot \dot{m}_x^{-\zeta-1} \right]^{-1} .
  \label{eq:DiskModel}
\end{equation}
To simplify, we define the constant $\zeta \equiv \xi - \gamma(2-\beta)$.  The
behavior is determined by the form of $\Sigma(R_h)$.  For $\dot{R}_h < 0$, the
surface density of the hot disk is relevant:  $\Sigma(R) = \Sigma_\hot(R)$
from \eq{HotDiskProfile}.  For $\dot{R}_h > 0$, $\Sigma(R)$ is the surface
density of the cold disk, which depends on the state of the disk prior to
outburst.  As the forms of $\Sigma_\hot$ and $\Sigma_\cold$ are generally
different, we must handle the two cases separately.

During the decay of an outburst from a partially ionized disk, the fall of
$\dot{M}_x$ will cause $R_h$ to move inward.  As it does, the surface density of
matter that this cooling front encounters will follow the quasi-stable hot
disk profile of \eq{HotDiskProfile}, and from assumption 2 this
$\Sigma_\hot(R_h)$ will become ``frozen'' into the cold disk profile as $R_h$
passes inward.

This ``freezing in'' of the hot disk surface density provides another test of
the hypothesis that the light curve knee marks the transition from a fully to
partially ionized disk.  The resulting density profile left behind is
\begin{equation}
  \Sigma(R) = \frac{\dot{M}_{x,c}}{3\pi\nu_0}
    \left(\frac{R}{R_\disk}\right)^{-\beta+(1-\xi)/\gamma} \, .
\end{equation}
For the $\gamma$, $\xi$, and $\beta$ of an irradiation-dominated disk,
\begin{equation}
  \Sigma(R) = 0.5\times\frac{\tau_\hot\dot{M}_{x,c}}{R_\disk^2}
    \left(\frac{R}{R_\disk}\right)^{1.7} \, .
  \label{eq:PostOutburstProfile}
\end{equation}
If we assume full ionization at $f_{x,c}$, then we can measure $\tau_\hot =
6.6$~d from the initial decay of the 2004 outburst \citep{Falanga05} and use
$f_x / f_{x,c}$ = $\dot{M}_x / \dot{M}_{x,c}$ to convert between flux and the
relative accretion rate.  Integrating \eq{PostOutburstProfile} over the disk
then gives a post-outburst mass sufficient to account for only 70\% of the
observed fluence during the 2008b outburst.  No physical viscosity model will
leave 95\% of the 2008b fluence remaining in the disk after the 2008a
outburst, as required.  Again, it is difficult to reconcile the observed
outbursts with the accretion disk of \igrj ever being fully ionized.

In contrast, this approach can successfully model the outburst light curve
when $f_x > f_{x,c}$ if we assume a partially ionized disk.  In this scenario
\eq{DiskModel} can be solved analytically, giving a decay for a partially
ionized disk with the same form as the fully ionized case but with $\zeta$ as
its shape parameter:
\begin{equation}
  \dot{M}_x = \left(\dot{M}_{x,0}^{-\zeta} + 
    \dot{M}_{x,c}^{-\zeta} \cdot \zeta\frac{t}{\tau_\hot}\right)^{-1/\zeta} \, .
  \label{eq:LateDecay}
\end{equation}
Assuming a constant viscosity and a disk scale height that increases linearly
with radius, we obtain $\zeta = -1$, giving the linear decay predicted by KR98
and P07.  For an irradiation-dominated disk and a gas-pressure-dominated
\citet{Shakura73} disk, the shape parameter is $\zeta = -1.17$ and $\zeta =
-1.30$, respectively.  The observed decay shape parameter of $-1.05\pm0.13$
when $f_x > f_{x,c}$ is compatible with the expected irradiated disk.
However, another mechanism must be invoked to explain the knee at $f_{x,c}$.

\subsection{Magnetospheric Inhibition of Accretion}
\label{sect:MTAndLC}

The drop in the accretion rate when the flux falls below $f_{x,c}$ may also be
caused by the onset of magnetospheric effects.  This explanation has been
invoked to explain similar drops in the light curves of other NS LMXBs (e.g.,
Aql X-1, \citealt{Campana98}; \saxj, \citealt{Gilfanov98}).  As the accretion
rate declines, the magnetospheric radius moves outward: $R_m \propto
\dot{M}_x^{-2/7}$.  When $R_m$ exceeds the co-rotation radius $R_\co$, at
which the Keplerian orbital frequency equals the NS spin frequency, infalling
matter must accelerate to co-rotate with the magnetic field.  At this point
the source is typically said to enter the ``propeller'' regime, as it was
originally though that centrifugal acceleration would eject matter from the
system \citep{Illarionov75}.  In fact, this picture is not energetically
self-consistent, and $R_m / R_\co \gtrsim 1.2$ is required for mass ejection
\citep{Rappaport04, Perna06}.  Below this limit, matter will build up in the
vicinity of $R_m$ until its material pressure pushes the magnetosphere inward,
allowing matter to accrete onto the NS and relieving the pressure on $R_m$
\citep{Spruit93}.  We refer to this intermediate state as the
``quasi-propeller'' accretion mode, because the centrifugal acceleration of
infalling matter throttles but does not entirely stop accretion.

Modeling the throttling of the accretion rate induced by the onset of a
quasi-propeller state is beyond the scope of this paper, but we consider some
qualitative predictions.  In particular, it can account for the different
shapes of the outburst light curves during 2004 and 2008a, which both exhibit
a fast rise and a linear-then-exponential decay, and 2008b, which has a
roughly linear rise, a plateau, then a roughly linear fall.  The key to this
difference is the distribution of matter in the accretion disk prior to the
outbursts.  Additionally, the rapid drop in flux caused by the onset of the
quasi-propeller state allows enough mass to remain in the disk following the
2008a outburst to fuel the 2008b outburst.

First, consider the 2004 and 2008a outbursts.  An accretion disk that has been
quiescent for a sufficient length of time will relax into a constant-density
state (KR98):
\begin{equation}
  \Sigma(R) \propto H \propto R^n \approx R^{1.2} \, .
\end{equation}
Distributing an initial disk mass accordingly and integrating \eq{DiskModel},
we get a light curve that rises faster than exponentially.  This rise breaks
our first assumption, that the hot disk can maintain a quasi-steady-state
density profile, but it is qualitatively instructive: it predicts the fast
rise observed for an outburst following a long quiescence.  For these
outbursts the rising flux never became bright enough to fully ionized the
accretion disk, causing the initial outburst decay to be approximately linear,
as discussed in the previous section.  Finally, when the flux reached
$f_{x,c}$ magnetospheric throttling becomes effective, causing the decay rate
to greatly increase and ultimately quenching the outburst.

The accretion disk that was present at the beginning of outburst 2008b would
not have had time to relax into a quiescent profile, so it depended on the
profile left behind by 2008a.  Consider the instantaneous state of the
accretion disk when the 2008a outburst reached the knee at $f_x = f_{x,c}$.
In the cold region of the disk (i.e., outside the $R_h$ corresponding to the
flux $f_{x,c}$), the disk profile left behind by the cooling front will follow
\eq{PostOutburstProfile}:
\begin{equation}
  \Sigma(R) \propto R^{-\beta+(1-\xi)/\gamma} \approx R^{1.7}
  \quad \textrm{for } R \ge R_h(f_{x,c}) \, .
  \label{eq:OuterDiskAtMTOnset}
\end{equation}
In the ionized region, the quasi-steady surface density profile of
\eq{HotDiskProfile} will be present:
\begin{equation}
  \Sigma(R) \approx \frac{\dot{M}_x}{3\pi\nu} \propto R^{-\beta} \approx R^{0.4}
  \quad\ \textrm{for } R \le R_h(f_{x,c}) \, .
  \label{eq:InnerDiskAtMTOnset}
\end{equation}
As the flux continued to fall, the onset of magnetospheric throttling caused
the decay time scale to becomes faster than the viscous time scale.  The
result is a greater amount of mass remaining in the inner disk than predicted
by \eq{PostOutburstProfile}: 40\% more matter would have been accreted if the
light curve had continued to follow a linear decay rather than becoming
exponential.  The distribution of the remaining inner disk mass will fall
between its state at the onset of magentic throttling and the distribution
left behind by freezing in the quasi-steady profile:  $0.4 < d\log\Sigma /
d\log R < 1.7$ for $R \le R_h(f_{x,c})$.  Finally, during the 30~d quiescence
the disk profile would begin to relax toward the constant-density $R^{1.2}$
distribution, but it likely would not have enough time to reach it.

This disk profile has two repercussions.  First, the faster shutoff of
accretion should leave sufficient mass in the disk to fuel the 2008b outburst.
Second, the distribution of this mass can result in the observed 2008b light
curve.  $\Sigma(R)$ of the inner disk has a smaller gradient than it would
after a long period of quiescence, so the amount of mass that $R_h$ encounters
as it expands outward increases more gradually.  As a result, we do not get
the sudden, super-exponential brightening seen for 2004 and 2008a.  Once the
flux reaches $f_{x,c}$, however, the heating front will encounter a lower
surface density, as the disk ionized by fluxes higher than $f_{x,c}$ will have
been more fully depleted by 2008a.  This change in $d\log\Sigma/d\log R$ will
stop the advance of $R_h$, initiating the outburst decay.  The result is a
slow rise, slow decay outburst that peaks at $f_{x,c}$, as observed.

We have not attempted to explain what stops the rapid rise of the 2004 and
2008a outbursts.  One possibility is self-shadowing by the accretion disk: a
warped disk or a point at which the radial dependence of the disk height
decreased would halt or slow the outward movement of the ionization front
during the outburst rise.  Regardless of the mechanism, the lower peak flux of
the 2008a outburst caused it to more quickly reach the onset of rapid decay at
$f_{x,c}$, ultimately leading to a shorter outburst with an accreted mass of
roughly half the mass consumed during the 2004 outburst.  It is likely that
the lower peak flux of 2008a was a necessary condition for enough matter to be
left in the disk to fuel the 2008b outburst.

\section{Evidence for Changes in the Accretion State}

The previous section showed how the transition of \igrj from steady accretion
to a quasi-propeller state at the light curve knee could successfully account
for the other features of the outburst light curves.  This explanation has two
requirements of the accretion rate corresponding to the knee: first, it must
not fully ionize the accretion disk; and second, it must be compatible with
the expected accretion rate at which the NS magnetosphere begins transferring
angular momentum to the infalling material.  In this section we address these
requirements.  We also consider other evidence suggesting that \igrj enters a
quasi-propeller state during its outbursts.  Throughout, we compare with the
401~Hz AMSP \saxj, which also shows light curve knees that are most likely
associated with the onset of the quasi-propeller state (e.g.,
\citealt{Hartman09a, Patruno09d}).

\subsection{The Accretion Rate at the Light Curve Knee}

Distance estimates to \igrj cover 2--5~kpc (see \citealt{Torres08} for a
recent review); we choose 4~kpc for our calculations, which is consistent with
mass transfer arguments and the lack of observed thermonuclear bursts
\citep{Galloway05}.  The bolometric correction factor for our 2.5--25~keV
fluxes is $c_\bol = 2.54$ \citep{Galloway05}.  With an assumption of a
canonical NS mass and radius ($M_x = 1.4\ M_\sun$, $R_x = 10$~km), the
accretion rate at the light curve knee is
\begin{equation}
  \label{eq:MdotKnee}
  \dot{M}_{x, c} = \frac{4\pi d^2 R_x}{GM_x} c_\bol f_{x,c} =
  1.8\times10^{-10}\ M_\sun\ {\rm yr}^{-1} \, .
\end{equation}
The distance uncertainty introduces a factor of $\sim$2 uncertainty in this
figure.

The outburst light curves of \saxj also show a knee at which decay steepens.
Assuming a distance of 3.5~kpc \citep{Galloway06} and identical NS parameters,
the accretion rate at that source's knee is also
$1.8\times10^{-10}\ M_\sun\ {\rm yr}^{-1}$ \citep{Hartman09a}.  The perfect
agreement of these numbers is a coincidence, of course, but it does strongly
suggest that the light curve knees of the two sources arise from the same
physical process.  Considerable evidence points to this knee marking the
beginning of the quasi-propeller state for \saxj.  A $\sim$1~Hz QPO with a
very high amplitude (sometimes $> 100\%$~rms) is present after the knee of
some outbursts.  \citet{Patruno09d} argue that a natural explanation is
provided by the \citet{Spruit93} instability, wherein matter that accumulates
near the magnetospheric radius quasi-periodically drips onto the NS surface.
\citet{Hartman09a} show that the behavior of the soft lags of the pulsations
also changes at this critical flux in a way that suggests a change in the
accretion column geometry that would occur when $R_m \approx R_\co$.

Similar evidence suggest that a quasi-propeller state turns on in \igrj at
fluxes below $f_{x,c}$.  A $\sim$0.5~Hz QPO was present throughout the 2008b
outburst, for which the flux was $\lesssim f_{x,c}$ at all times.  Although
its 13\%~rms amplitude was much less than the amplitude of the 1~Hz QPO of
\saxj, it is plausible that it too is due to the Spruit-Taam instability.  The
energy dependence of the \igrj pulsations remained constant through the
outbursts, so it offered no evidence either way.  We will consider other
effects indicating a strong disk-magnetosphere interaction later in this
section.

The hypothesis that the accretion disk of \igrj is never fully ionized during
the observed outbursts is perhaps more controversial.  In
section~\ref{sect:IonizedDiskDecay}, we argued that the observed linear decay
during the first 7~d of the 2004 ruled out a fully ionized disk.  The same is
true for \saxj: we find that the high-flux, slow-decay stages of its 1998,
2005, and 2008 outburst light curves are better fit with linear models than
exponential models,\footnote{The 2002 outburst of \saxj, which had a peak flux
  50\% brighter than the other outbursts, initially followed an approximately
  exponential decay until reaching a first knee while still at a relatively
  high flux.  The decay then slowed, but whether its subsequent form was
  better fit with an exponential or linear model depends on the region fit.
  After a second knee at that source's critical flux, it entering rapid decay
  (see Fig.~3 of P07 and Fig.~1 of \citealt{Hartman08}).  The peak flux of
  this outburst possibly does fully ionize the accretion disk, with the first
  knee marking the transition to partial ionization.} while the rapid decay
following the knees of its outbursts are approximately exponential (e.g.,
\citealt{Gilfanov98}).  Aql~X-1 shows a similar transition to exponential
decay as it fades \citep{Campana98}.  On the other hand, the calculations of
KR98 predict that short-period transients should fully ionize their disks, and
full ionization has been invoked to account for the light curves of many NS
LMXBs (e.g., \citealt{Shahbaz98}; P07).

The 2004 outburst had a peak flux of $2.5\ f_{x,c}$, implying an accretion
rate of $4.5\times10^{-10}\ M_\sun$~yr$^{-1}$.  KR98 predicts that the entire
disk will be ionized above a critical rate of
\begin{equation}
  \dot{M}_{x,{\rm KR}} = 4.1\times10^{-10}
    \left(\frac{R_\disk}{10^6\ {\rm km}}\right)^2\ M_\sun\ {\rm yr}^{-1}
  \, .
\end{equation}
The disk radius of \igrj must be smaller than $10^6$~km for the observed
orbital parameters and realistic masses, causing an apparent contradiction.
That said, the disk model used by KR98 to derive the above critical rate is
quite basic and likely requires substantial modification to produce a physical
disk structure in a tight binary.  A significant shortcoming is the assumption
of a constant $n = d\log H / d\log R$ to describe the vertical profile.
Simulations of irradiated disks produce profiles that flatten out at larger
radii (e.g., \citealt{Dubus99}), screening the outer disk from irradiation and
increasing the $\dot{M}_x$ required to ionized the non-screened portion of the
disk.  Additionally, irradiation will tend to warp LMXB accretion disks
\citep{Pringle96}, allowing irradiation of some outer regions while shadowing
others.  The profile of an irradiated disk remains an open question, but it
most likely can accommodate a partially ionized disk even at high accretion
rates.

Finally, we must consider the detection of \Halpha emission during the 2004
and 2008 outbursts \citep{Torres08, Lewis10} under the assumption of partial
ionization.  The \Halpha lines were double-peaked and showed no variation with
orbital phase, clearly indicating their origin in the disk.  \citet{Torres08}
reported a peak-to-peak separation of $\Delta v_{\rm pp} =
650\pm40$~km~s$^{-1}$ on 2004 Dec~5 (MJD 53344), near the peak of the 2004
outburst.  Assuming that this $\Delta v_{\rm pp}$ measured the line-of-sight
velocity at the edge of a fully ionized accretion disk, they estimated an
orbital separation of $a = (7.2$--$8.3)\times10^5$~km, an inclination of $i =
22\degr$--32\degr, and a companion mass of $M_2 = 0.07$--0.11~$M_\sun$ for a
1.4~$M_\sun$ NS (0.09--0.13~$M_\sun$ for a 2.0~$M_\sun$ NS).  If the
assumption of full ionization is dropped, then their calculated $a$ and $M_2$
ranges become lower limits, and their $i$ becomes an upper limit.  A weaker
upper limit on $M_2$ is provided by the requirement that the companion fits
within its Roche lobe, giving $M_2 \lesssim 0.25\ M_\sun$ \citep{Galloway05,
  Torres08}.  The resulting range of allowed inclinations is 10\degr--32\degr.
A low inclination for this system is supported by the lack of harmonic content
in the X-ray pulsations (e.g., \citealt{Muno02}).

\subsection{The Magnetospheric Radius at the Knee}

As discussed at the beginning of Section~\ref{sect:MTAndLC}, the mass
accretion (or ejection) regime depends on the magnetospheric radius,
\begin{equation}
  R_m = k_m (2 G M_x)^{-1/7} \dot{M}_x^{-2/7} \mu^{4/7} \, .
  \label{eq:R_m}
\end{equation}
Here $\mu$ is the magnetic dipole moment, and $k_m$ is an order-unity constant
that encapsulates the complex disk/magnetosphere interaction.
Magnetohydrodynamic simulations of \cite{Long05} suggest a value of $k_m
\approx 0.5$, which we adopt for our calculations.  The accretion mode depends
on the relation between $R_m$ and the Keplerian co-rotation radius $R_\co =
(GM_x / 4\pi^2\nu^2)^{1/3}$.  Roughly speaking, for $R_m \lesssim R_\co$,
matter can accrete steadily onto the star; for $R_m \approx R_\co$,
centrifugal acceleration of infalling matter by the NS magnetosphere slows but
does not entirely inhibit accretion, a regime we refer to as the
quasi-propeller mode; and for $R_m \gtrsim R_\co$, a true propeller effect as
described by \citet{Illarionov75} becomes energetically possible, ejecting
matter from the system and thereby preventing accretion.  For AMSPs, the
transition zone around $R_m$ from a Keplerian to a co-rotating flow is wide,
so changes in the mode of accretion are likely to be gradual.

Magnetohydrodynamic simulation, analytic models, and observational evidence
all indicate that the transition from steady accretion to a quasi-propeller
regime occurs around $R_m \approx 0.7\ R_\co$.  The magnetohydrodynamic
simulations of \citet{Long05} identified this ratio as the critical
magnetospheric radius below which $\dot\nu > 0$ and above which $\dot\nu < 0$.
The analytic model of \citet{Rappaport04} produces a similar threshold between
spin-up and spin-down.\footnote{The ratio given in \citet{Rappaport04} is $R_m
  \approx 1.87\ R_\co$, but this relation is based on a formal definition of
  $R_m$ that is a factor of 2.2 greater than ours.  Adjusting for this
  difference gives $R_m \approx 0.85\ R_\co$.}  Finally, the well-constrained
distance and magnetic field of \saxj allow us to calculate its $R_m$ from its
observed flux using \eq{R_m}.  The knee in the decay of that source (along
with other phenomena indicating the onset of a quasi-propeller state) occurs
when $R_m = 0.6\ R_\co$ \citep{Hartman09a}.

The magnetospheric radius implied at the knee of the \igrj outburst light
curves is also consistent with this value.  The long-term quiescent spin-down
of this source suggests a magnetic dipole moment of $\mu =
1.1\times10^{26}$~G~cm$^2$, which we derive in
section~\ref{sect:DipoleSpindown}.  From this field strength and the critical
$\dot{M}_x$ derived in \eq{MdotKnee}, \eq{R_m} gives us
\begin{equation}
  \frac{R_{m,{\rm knee}}}{R_\co} = \frac{16\ {\rm km}}{23.6\ {\rm km}}
   = 0.7 \, .
\end{equation}
This figure is in excellent agreement with theoretical expectations and our
observations of \saxj.  It further reinforces the validity of our assumption
that the light curve knee marks the transition into a quasi-propeller
accretion state.

\subsection{Halting of Accretion by the Propeller Effect?}

The transition to a true propeller state, in which the centrifugal
acceleration is capable of ejecting matter from the system, occurs at $R_m
\approx 1.26\ R_\co$ if the transfer of angular momentum is perfectly
efficient but inelastic \citep{Rappaport04}; if the transfer is also perfectly
elastic, this limit becomes $R_m \approx 1.13\ R_\co$ \citep{Perna06}.  From
\eq{R_m}, these radii respectively correspond to $f_x/f_{x,c} \approx 0.12$
and 0.19, giving 2.5--25~keV fluxes of $(5$--$8)\times10^{-11}$\fluxunits for
\igrj.  Within or slightly below this range, we predict another knee in the
light curve as the decay rate again increases and the outburst rapidly shuts
off.

\begin{figure}[t!]
  \begin{center}
    \includegraphics{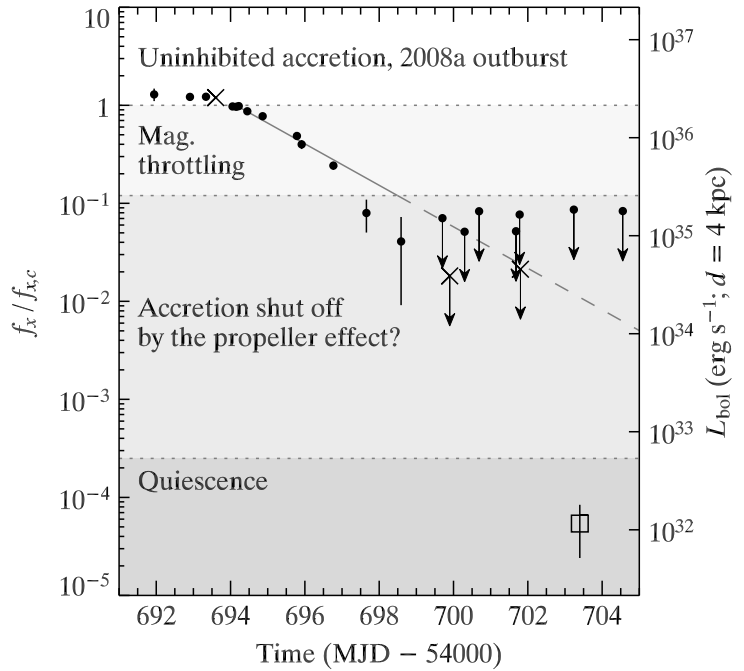}
  \end{center}
  \caption{ A comparison of the observed fluxes and predicted accretion
    regimes during the 2008a outburst.  Fluxes are from the PCA (dots), \Swift
    XRT ($\times$ marks), and \XMM (square).  (\Swift and \XMM measurements
    are taken from \citealt{Lewis10}.)  Horizontal dotted lines indicate the
    onset of magnetic throttling ($f_x/f_{x,c} = 1$ by assumption), the
    approximate flux below which the propeller effect should halt accretion
    entirely ($f_x/f_{x,c} \approx 0.12$), and the typical flux observed by
    \Chandra during quiescence (see text for references).  The solid gray line
    shows an exponential fit to the light curve during the magnetic throttling
    regime, and the dashed line extrapolates this fit to later times.  Note
    that the \XMM flux falls far below this trend line.  There must be another
    break in the light curve around the time that the source falls below the
    PCA detection level.
  \label{fig:AccRegimes}}
  \vspace{1em}
\end{figure}

\Swift XRT and \XMM observations following the 2008a outburst require the
presence of such a break.  The trend line in Figure~\ref{fig:AccRegimes} shows
the exponential decay fit to the PCA fluxes in the range $1 \ge f_x/f_{x,c}
\ge 0.12$, for which the NS magnetosphere should slow but not entirely halt
accretion.  Its $e$-folding time is $1.8\pm0.3$~d, as reported in
section~\ref{sect:LightCurveResults}.  After the outburst falls below
$f_x/f_{x,c} \approx 0.2$, the decay accelerates.  The two last \RXTE
observations in which \igrj is confidently detected fall somewhat (2--3\sig)
below the extrapolated exponential decay curve.  1.4~d after the last \RXTE
detection, a 3\sig \Swift XRT upper limit falls well below it \citep{Lewis10}.
Finally, a deep \XMM observation 11~d after the outburst peak and 5~d after
the last \RXTE detection revealed that \igrj had dropped to an unabsorbed
2--10~keV flux\footnote{Using the spectral models of \S\ref{sect:SpecParams},
  we calculate a factor of 1.7 to convert from the unabsorbed 2--10~keV fluxes
  given by \citet{Lewis10} to the absorbed 2.5--25~keV fluxes quoted in this
  paper.} of $1.4\times10^{-14}$\fluxunits.  This flux is lower than
previously reported quiescent levels by a factor of $\sim$3 \citep{Jonker05,
  Jonker08, Torres08}, although \citet{Jonker05} notes that the quiescent
X-ray flux can vary.  \igrj could not have reached this quiescent flux so
quickly without another steepening of the decay: fitting a constant decay rate
between the last PCA detections and the \XMM observation establishes a maximum
$e$-folding time of 0.7~d.  The actual decay rate was likely faster.

If the majority of optical emission during outburst is due to X-ray
reprocessing, then we should expect a knee in the optical light curve if and
when the mass accretion rate becomes low enough to be entirely halted by the
propeller effect.  Such a break was seen in the R band during the 2004
outburst when the X-ray flux was $f_x/f_{x,c} = 0.13\pm0.03$, in excellent
agreement with our predictions \citep{Torres08}.  Gaps in the optical coverage
of the 2008a and 2008b outbursts around the expected times of breaks prevented
their detection or exclusion (\citealt{Lewis10}; F. Lewis \& D. M. Russell,
priv. comm.).

\subsection{Accretion Torques during Outburst}
\label{sect:OutburstTorques}

Changes in the accretion state should also change the X-ray timing properties
of \igrj.  At accretion rates above the onset of the quasi-propeller state,
the infalling matter should spin up the NS.  When accretion enters the
quasi-propeller state, which we suggest corresponds to the knee in the X-ray
light curve, the transfer of angular momentum to the NS becomes inefficient or
negative.

The 2004 outburst of \igrj, which had a peak flux of roughly twice what was
seen during 2008, showed a clear spin-up of $8.4(6)\times10^{-13}$\HzPerSec,
which \citep{Burderi07} attributed to accretion torque.  Under the assumption
that the light curve knee approximately corresponds to the spin-up / spin-down
equilibrium point, this result is expected.

In contrast, the $\dot\nu$ measurements for the 2008 outbursts were poorly
constrained:  $10(10)\times10^{-13}$\HzPerSec and
$4.5(2.5)\times10^{-13}$\HzPerSec respectively.  At 1.0\sig and 1.8\sig
significance, it is possible that these marginal spin-ups reflect timing noise
rather than any change in the spin of the NS.  If they are real, the frequency
at the end of the 2008a outburst would be 0.5(2)~\uHz higher than the
frequency at the beginning of the 2008b outburst, requiring a mean spin-down
of $-1.7(6)\times10^{-13}$\HzPerSec during the intervening 30~d quiescence.
Spin-ups during the 2008 outbursts would also increase the implied long-term
spin-down.

Yet the 2008b outburst, which never rose above $f_{x,c}$, was likely spent
entirely in the quasi-propeller state or at the limit of its onset.  We should
therefore expect the NS to be spun down during this outburst.  This is at odds
with its 1.8\sig spin-up measurement.  Given the low significance of this
detection and the otherwise solid evidence that the knee at $f_{x,c}$ reflects
a transition into a quasi-propeller state, it is likely that this $\dot\nu$ is
due to timing noise rather than a change in the NS spin.  The case of the
2008a outburst is similar: because it spent little time ($\approx$2.5~d) above
$f_{x,c}$, a large spin-up is not expected.

\section{Magnetic Field Limits from Pulsations}

The presence of accretion-powered pulsations across a nearly two orders of
magnitude in flux constrains the magnetic field strength of the NS.
\citet{Psaltis99} used the similarly wide range of fluxes with detectable
pulsations from \saxj to derive limits that were compatible with the magnetic
field implied by the spin-down of that source \citep{Hartman09b}.  Here we
apply their arguments to \igrj.

The detection of pulsations when \igrj was at its peak flux indicates that the
magnetic field must be strong enough to columnate the accretion flow above the
NS surface even when the accretion rate is at its maximum.  Setting the
magnetospheric radius $R_m$ from \eq{R_m} equal to the NS radius $R_x$ and
solving for the magnetic dipole moment $\mu$ gives a lower limit:
\begin{eqnarray}
  \mu & \ > \ & 1.6\times10^{25} \mathrm{\ G\ cm^3\ }
  \left(\frac{k_m}{1.0}\right)^{-7/4} \nonumber\\ & & \times
  \left(\frac{M_x}{2.3\ M_\sun}\right)^{-1/4}
  \left(\frac{R_x}{10\ {\rm km}}\right)^{9/4}
  \left(\frac{d}{3\ {\rm kpc}}\right) \nonumber\\ & & \times
  \left(\frac{c_\bol \cdot f_{x,{\rm max}}}
    {2.54 \cdot 11.1\times10^{-10}\mathrm{\ erg\ cm^{-2}\ s^{-1}}}\right)^{1/2}
\end{eqnarray}
For this conservative limit, we have taken extreme values of the
magnetospheric constant $k_m$, the NS mass and radius, and the distance to the
source.

Conversely, at the lowest fluxes with detectable pulsations, the magnetic
field cannot be so strong that it causes the ejection of matter due to the
propeller effect.  This propeller regime turns on at around $R_m = 1.3\ R_\co$
\citep{Rappaport04}.  Solving for $\mu$, we get an upper limit:
\begin{eqnarray}
  \mu & \ < \ & 1.2\times10^{27} \mathrm{\ G\ cm^3\ }
  \left(\frac{k_m}{0.2}\right)^{-7/4}
  \left(\frac{\nu}{599\ {\rm Hz}}\right)^{-7/6} \nonumber\\ & & \times
  \left(\frac{M_x}{2.3\ M_\sun}\right)^{1/3}
  \left(\frac{R_x}{15\ {\rm km}}\right)^{1/2}
  \left(\frac{d}{5\ {\rm kpc}}\right) \nonumber\\ & & \times
  \left(\frac{c_\bol \cdot f_{x,{\rm min}}}
    {2.54 \cdot 0.66\times10^{-10}\mathrm{\ erg\ cm^{-2}\ s^{-1}}}\right)^{1/2}
\end{eqnarray}
Again, we have adopted parameter values that give a conservative estimate of
this limit.

\section{Implications of the Long-term Spin-down}
\label{sect:LongTermSpinDown}

The measured 2004 and 2008 frequencies give a mean long-term spin-down of
$-(4\pm1)\times10^{-15}$\HzPerSec during the 3.7~yr quiescence.  This large
$\dot\nu$ range is principally due to the uncertainty of whether the 2008
outburst spin-ups are real.  If they are not, as suggested by
Section~\ref{sect:OutburstTorques}, then the long-term spin-down is at the
bottom of this range:  $-(2.8\pm0.5)\times10^{-15}$\HzPerSec.  For the rest of
this section, we will use this more conservative value.  Adopting the higher
figure would only strengthen our conclusions here.

The resulting spin-down luminosity is $\dot{E} = -4\pi^2 I \nu \dot\nu =
7\times10^{34}$~erg~s$^{-1}$.  This figure is eight times greater than the
spin-down luminosity of \saxj \citep{Hartman09b}, due to both $\nu$ and
$\dot\nu$ being higher for \igrj.  We consider three possible mechanisms for
this spin-down.

\subsection{The Propeller Effect}

In section~\ref{sect:MTAndLC}, we discussed the centrifugal expulsion of
matter by the NS magnetosphere, commonly known as the propeller effect.  This
ejection of matter will produce a spin-down of the NS; however, it is unlikely
that it is sufficient to explain the large frequency change observed between
the 2004 and 2008 outbursts.

The propeller effect will eject matter from near the magnetospheric radius,
producing a torque of
\begin{equation}
  N_{\rm prop} \approx \dot{M}_{\rm ej} (G M_x R_m)^{1/2}
\end{equation}
if matter is ejected at a rate of $\dot{M}_{\rm ej}$.  Centrifugal ejection
occurs if $R_m \gtrsim 1.3\ R_\co \approx 30$~km.  $R_m$ varies weakly with
$\dot{M}_{\rm ej}$, so it will not be much greater than this value.  The
$\Delta\nu = 0.32(4)$~\uHz frequency difference between the end of the 2004
outburst and the beginning of 2008a would then require a total mass ejection
of $\sim$$4\times10^{-11}\ M_\sun$ during the 3.7~yr quiescence.  The
associated rate is an order of magnitude greater than the expected mass
transfer rate of $3\times10^{-12}\ M_\sun$~yr$^{-1}$ from the low-mass
companion assuming that transfer is driven by gravitational radiation from the
binary orbit \citep{Galloway05}.

\begin{figure}[t!]
  \begin{center}
    \includegraphics[width=0.45\textwidth]{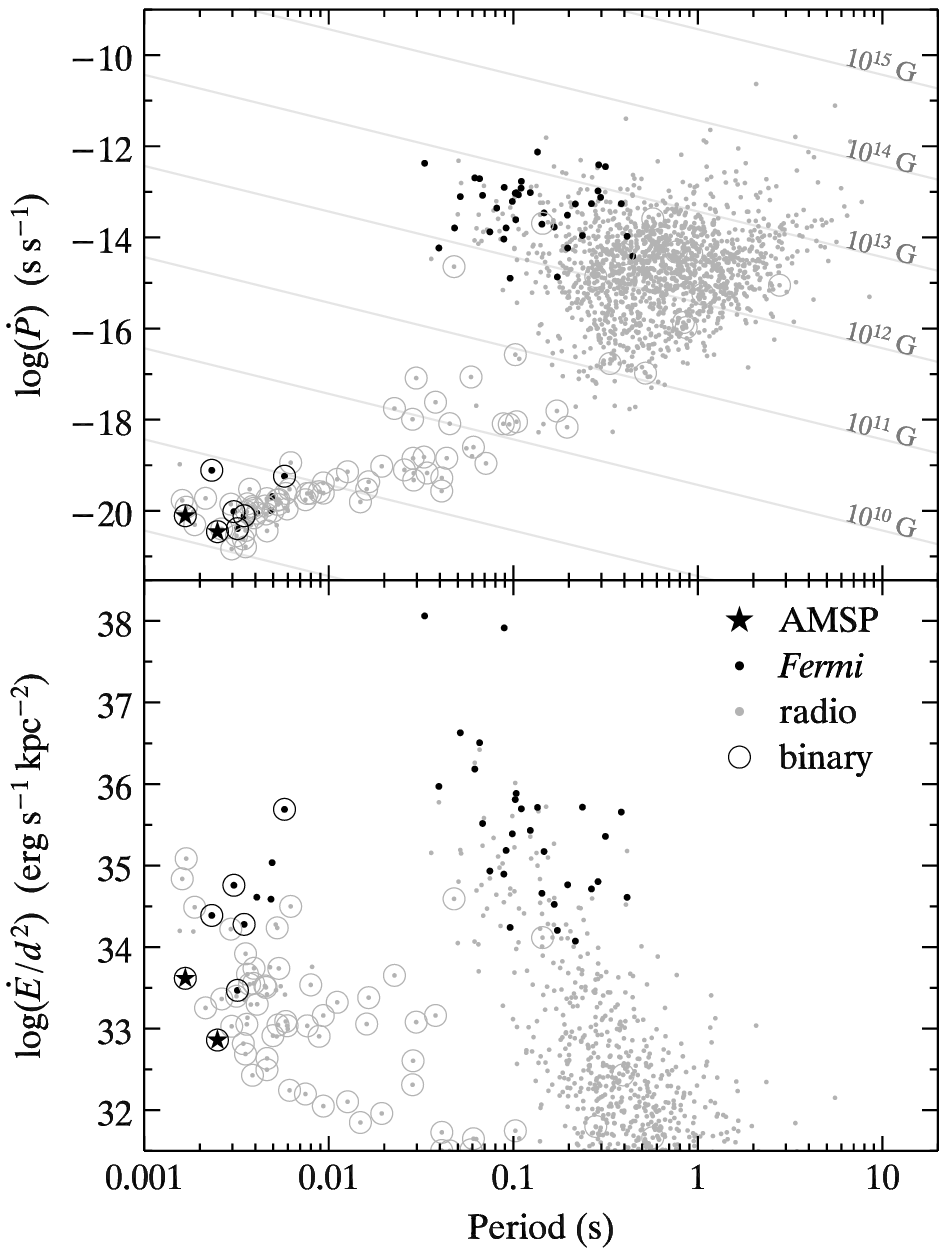}
  \end{center}
  \caption{ A comparison of the AMSPs (stars), \Fermi pulsars (black points),
    and radio pulsars without \gammaray emission (gray points).  Pulsars in
    binary systems are circled.  The top plot shows the standard $P$-$\dot{P}$
    diagram.  The two AMSPs with known quiescent spin-downs, \saxj and \igrj,
    are located in the lower left corner, among the radio millisecond pulsars.
    Lines of constant magnetic field were calculated using the formula of
    \citet{Spitkovsky06}.  The bottom plot shows the spin-down luminosities,
    normalized by distance squared.  These values are proportional to the
    spin-down flux that would be incident if all the sources radiated with
    perfect efficiency.  Fermi pulsar parameters are from \citet{FermiPsrCat};
    radio pulsar parameters are from the Australia Telescope National Facility
    pulsar catalog \citep{ATNFPsrCat}, online at {\tt
      http://www.atnf.csiro.au/research/pulsar/psrcat}.  Pulsars in globular
    clusters have been omitted.
  \label{fig:PulsarDiagrams}}
\end{figure}

\subsection{Magnetic Dipole Spin-down}
\label{sect:DipoleSpindown}

Magnetic dipole spin-down is a more likely cause.  Pulsar magnetosphere
simulations by \citet{Spitkovsky06} give a magnetic dipole torque of $N =
-\mu^2(2\pi\nu/c)^3(1+\sin^2\alpha)$.  Here $\alpha$ is the colatitude of the
magnetic pole, which we assume to be small ($\approx$15\degr) based on the
highly sinusoidal pulse profile (e.g., \citealt{Poutanen06, Lamb09}).  If this
torque accounts for the entirety of the quiescent spin-down, it would require
a magnetic dipole of $(9.4\pm0.8)\times10^{25}$~G~cm$^3$.  For a NS radius of
10~km, the corresponding surface field is $B = 2\mu R_x^{-3} =
1.9\times10^8$~G at the magnetic poles.  This field strength is consistent
with the limits derived in the previous section from magnetospheric arguments.
It also agrees with the $B < 3\times10^8$~G limit of \citet{Torres08} based on
the quiescent luminosity.  Finally, the long-term spin-down of \igrj places it
among the radio millisecond pulsars on the $P$-$\dot{P}$ diagram, suggesting a
common spin-down mechanism.

The high spin-down luminosity of \igrj makes it a good candidate for pulsation
searches during quiescence.  The discovery of spin-powered pulsations from a
quiescent AMSP would provide a final and elusive missing link between the
AMSPs and radio millisecond pulsars.  \gammaray pulsation searches may be even
more promising.  The first eight millisecond \gammaray pulsars detected by the
\Fermi Large Area Telescope (LAT) convert their spin-down luminosities into
\gammaray emission at high efficiencies (6--100\%) and emit pulsations into
wide opening angles \citep{FermiMSPs09}.  If \Fermi does not detect pulsations
from \igrj, it would require a lower dipole spin-down to \gammaray emission
efficiency from this source, or a different spin-down mechanism entirely.

\subsection{Gravitational Radiation Torque}

Angular momentum loss through gravitational wave emission has been suggested
as a way to explain the absence of very rapidly spinning ($\gtrsim$730~Hz)
millisecond pulsars \citep{Bildsten98, Chakrabarty03, Chakrabarty05}.  The
highly nonlinear $\sim$$\nu^5$ dependence of the gravitational wave torque on
spin rate $\nu$ means that this torque would dominate at the very highest spin
rates but be negligible for slower spins, with a rather sharp transition.
While there is no evidence that this mechanism is important at spins as slow
as 400~Hz \citep{Hartman08}, \igrj is a strong candidate since it is the most
rapidly spinning AMSP.  However, as in the case of \saxj \citep{Hartman08}, we
have already found that magnetic dipole torques due to the known magnetic
field strength of the pulsar likely accounts for most of the spin-down in
quiescence, suggesting that gravitational wave torques are unimportant even at
599~Hz.

For \igrj, the quiescent spin-down places an upper limit on the neutron star's
mass quadrupole moment of\footnote{The coefficient of \eq{qlimit} differs from
  the value derived using eq.~(17) of \citet{Hartman08}.  The coefficient in
  that paper is incorrect: it should read $1.4\times10^{36}$~g~cm$^2$, not
  $4.4\times10^{36}$~g~cm$^2$.}
\begin{eqnarray}
  Q & \ < \ & 1.2\times10^{36}
    \left(\frac{I}{10^{45}{\rm\ g\ cm}^{2}}\right)^{1/2}
    \left(\frac{\nu}{599 {\rm\ Hz}}\right)^{-5/2} \nonumber\\ && \times
    \left(\frac{-\dot\nu}{2.8\times10^{-15}\ {\rm Hz\ s^{-1}}}\right)^{1/2}
    \textrm{ g cm}^2 \, , \label{eq:qlimit}
\end{eqnarray}
or an ellipticity of $Q/I \lesssim 10^{-9}$ for moment of inertia $I$.  The
rapid spin of \igrj makes this limit approximately an order of magnitude more
stringent than the similarly derived limit for \saxj \citep{Hartman08}.  It is
appreciably lower than the upper limits on NS ellipticity from direct
gravitational wave detectors: targetted LIGO searches of radio millisecond
pulsars give a strongest upper limit of $Q/I < 7\times10^{-8}$ for the nearby
pulsar PSR~J2124$-$3358 \citep{Abbott10}.

The spin-down of \igrj begins to test predictions of the expected quadrupole
moments of accreting neutron stars.  Accretion could result in a mass
quadrupole through a variety of mechanisms.  Electron capture due to higher
temperatures base of the accretion columns \citep{Bildsten98, Ushomirsky00},
hydrodynamic turbulence due to accretion-induced differential rotation
\citep{Melatos10}, and the accumulation of ``mountains'' of accreted material
(e.g., \citealt{Haskell06}) all could produce ellipticities of up to $Q/I \sim
10^{-8}$; magnetic confinement of the accreted material could allow an even
higher limit \citep{Melatos05, Payne06}.  The absence of such a high
ellipticity in \igrj does not rule out any of these mechanisms, as their
predicitions depend on unknown NS parameters (e.g., the maximum strain upheld
by the NS crust, or the ohmic diffusion time scale governing the settling of
accreted material).  Nevertheless, we may be entering the regime in which it
is possible to begin constraining some of these parameters.

This upper limit for \igrj still does not exclude the possibility that
gravitational wave torques affect the fastest pulsar spins, since the torque
near the limiting $\approx$730~Hz spin frequency would be 2.7~times greater
than in \igrj for the same $Q$.  It is also feasible that gravitational wave
emission is present only for a short time in these sources, acting only during
and shortly after an outburst.  Our limits in this paper and in
\citet{Hartman08} are valid only for the magnitude of a persistent quadrupole
and do not exclude a larger quadrupole that dissipates substantially faster
than the $\sim$3~yr outburst recurrence period.

\section{Conclusions}

We have presented a comprehensive analysis of the outbursts of \igrj and a
self-consistent explanation for the observed behavior.  In contrast with
previous analyses of irradiated accretion disks in NS LMXBs, we posit that the
irradiation never fully ionizes the disk of this source during the outbursts
observed with \RXTE in 2004 and 2008.  Instead, we suggest that the NS
magnetic field and its interaction with the accretion disk plays a central
role in shaping the light curves of these outbursts.

The long-term spin-down of \igrj is most likely due to the magnetic dipole
torque of a $2\times10^8$~G field at the NS surface.  A magnetic field of this
strength will impede the infall of matter when the accretion rate falls below
$2\times10^{-10}\ M_\sun$~yr$^{-1}$, in excellent agreement with the critical
flux at which knees are observed in the 2004 and 2008a light curves. If these
knees are due to disk-magnetosphere interaction, then it is unlikely that the
disk was ever fully ionized: irradiation is not sufficient to ionize the
entire disk at the lowest accretion rates observed with \RXTE, and a second
knee would likely be detectable if the transition from fully to partially
ionized had occurred.  Futhermore, the near-linear decay observed during the
first week of the 2004 outburst is inconsistent with the exponential decay
expected for a fully ionized disk but matches well with theoretical
predictions when partial ionization is assumed.

Substantial evidence points to the onset of a ``quasi-propeller'' state at
accretion rates below the light curve knee.  By shutting off accretion more
quickly than if the outburst was driven solely by irradiation, enough matter
is left in the disk after the 2008a outburst to fuel the 2008b outburst 30~d
later.  Additionally, the mass distribution left behind in the disk differs
from the distribution expected after a long period of quiescence, priming the
disk for the slow rise and maximum at $f_{x,c}$ seen in the 2008b outburst.
In the quasi-propeller regime below the knee, accretion torques should be
small or negative, and the non-detection of spin derivatives during the 2008
outbursts stands in contrast with the significant spin-up seen during the
brighter 2004 outburst.  Accretion instabilities associated with the magnetic
throttling may explain the $\sim$0.5~Hz QPO observed during the 2008b
outburst.  Finally, if the light curve knee is associated with the onset of
the quasi-propeller effect, then there should be another knee near the \RXTE
PCA detection threshold beyond which the propeller effect can entirely eject
infalling matter from the system, entirely halting accretion; indeed, \Swift
XRT observations and a deep \XMM observation show that another light curve
break must be present, causing the source to return to quiescent levels within
days of falling below the \RXTE detection threshold.  Clearly there remains
much work to be done, however: our invocation of magnetospheric effects has
been largely qualitative or based on simplistic models, and more substantial
analysis and modeling is necessary to gain a fuller understanding.

\acknowledgements

We thank Jean Swank and the \RXTE team for scheduling these observations.  We
thank Diego Altamirano, Craig Heinke, Miriam Krauss, Fraser Lewis, Paul Ray,
Michael Rupen, Dave Russell, and Kent Wood for helpful conversations and an
anonymous referee for useful comments.  JMH thanks the NRAO for its
hospitality and the use of its facilities during the writing of this paper.

Two other papers analyzing the X-ray timing of these outbursts were posted on
arXiv.org almost concurrently with this one: \citet{Patruno10b} and
\citet{Papitto10}.  These authors derived compatible results for the spin
frequencies during the outbursts and the long-term spin-down between 2004 and
2008, although \citet{Patruno10b} claimed a higher significance for the
spin-ups during outburst.  We thank Alessandro Patruno and Alessandro Papitto
for useful discussions and for their comments on this paper.


\end{document}